\documentclass[a4paper,supercite]{article}
\usepackage{amsmath}
\usepackage{amssymb}
\usepackage{epsfig}
\usepackage{color}
\usepackage{float}
\usepackage{dcolumn}
\usepackage{booktabs}

\newfont{\rms}{cmr10 at 10pt}

\newcommand{\erfc}{{\mbox{erfc}}}
\newcommand{\infd}{{\mbox{d}}}

\newcommand{\II}{{\mathbb{I}}}

\newcommand{\ZZ}{{\mathbb{Z}}}

\newcommand{\VECk}{{\boldsymbol{k}}}

\newcommand{\VECm}{{\boldsymbol{m}}}

\newcommand{\VECr}{{\boldsymbol{r}}}

\newcommand{\VECF}{{\boldsymbol{F}}}

\newcommand{\VECxi}{{\mbox{\boldmath$\xi$}}}

\newcommand{\VECzero}{{\boldsymbol{0}}}

\newcommand{\oscoeff}{\widehat{p}}
\newcommand{\rB}{{\operatorname{B}}}
\newcommand{\rC}{{\operatorname{C}}}

\newcommand{\rD}{{\operatorname{D}}}

\newcommand{\rLJ}{{\operatorname{LJ}}}

\newcommand{\rM}{{\operatorname{M}}}
\newcommand{\rMD}{{\operatorname{MD}}}

\newcommand{\PB}{{\operatorname{PB}}}

\newcommand{\rTr}{{\operatorname{Tr}}}

\newcommand{\rb}{{\operatorname{b}}}
\newcommand{\rca}{{\operatorname{ca}}}

\newcommand{\rcut}{{\operatorname{cut}}}
\newcommand{\re}{{\operatorname{e}}}
\newcommand{\eff}{{\operatorname{eff}}}

\newcommand{\rfit}{{\operatorname{fit}}}

\newcommand{\rig}{{\operatorname{ig}}}
\newcommand{\rl}{{\operatorname{l}}}

\newcommand{\rr}{{\operatorname{r}}}
\newcommand{\rrod}{{\operatorname{rod}}}
\newcommand{\rs}{{\operatorname{s}}}
\newcommand{\rsr}{{\operatorname{sr}}}

\newcommand{\sfp}{{\text{\sf p}}}

\newcolumntype{d}[1]{D{.}{.}{#1}}

\newcommand{\D}{\displaystyle}

\begin{document}

\thispagestyle{empty}

\begin{center} \LARGE 
{\bf Theory and simulations of rigid polyelectrolytes}\\[1cm]
\end{center}
\begin{center}\large 
Markus Deserno\footnote{email: {\tt markus@chem.ucla.edu}}$^a$,%
Christian Holm\footnote{email: {\tt holm@mpip-mainz.mpg.de}}$^b$ \\[1cm]
{\it $^a$Department of Chemistry and Biochemistry, UCLA, USA}\\
{\it $^b$Max-Planck-Institut f\"ur Polymerforschung, Ackermannweg 10, 55128 Mainz, Germany}
\end{center}

\begin{abstract}
  We present theoretical and numerical studies on stiff, linear
  polyelectrolytes within the framework of the cell model.  We first
  review analytical results obtained on a mean-field Poisson-Boltzmann
  level, and then use molecular dynamics simulations to show, under
  which circumstances these fail quantitatively and qualitatively. For
  the hexagonally packed nematic phase of the polyelectrolytes we
  compute the osmotic coefficient as a function of density. In the
  presence of multivalent counterions it can become negative, leading
  to effective attractions. We show that this results from a reduced
  contribution of the virial part to the pressure.  We compute the
  osmotic coefficient and ionic distribution functions from
  Poisson-Boltzmann theory with and without a recently proposed
  correlation correction, and also simulation results for the case of
  poly(para-phenylene) and compare it to recently obtained
  experimental data on this stiff polyelectrolyte. We also investigate
  ion-ion correlations in the strong coupling regime, and compare them
  to predictions of the recently advocated Wigner crystal theories.
\end{abstract}

\newpage


\section{Introduction}
``Polyelectrolytes are polymers bearing ionizable groups, which, in
polar solvents, can dissociate into charged polymer chains (macroions)
and small counterions'' \cite{barrat96a}. The combination of
macromolecular properties and long-range electrostatic interactions
produces an impressive variety of phenomena
\cite{hara93a,dautzenberg94a,foerster95a}. It makes these systems
interesting from a fundamental as well as a technological point of
view.

A thorough understanding of polyelectrolytes has become increasingly
important in biochemistry and molecular biology. This is due to the
fact that virtually all proteins, as well as the DNA, are
polyelectrolytes.  Their interactions with each other and with the
charged cell-membrane are still far from being understood. For
instance, a puzzling question is why two equally charged objects
should attract each other in the first place \cite{gelbart00a}.

In this article we focus on stiff linear polyelectrolytes, which we
subsequently approximate by charged cylinders. This is a relevant
special case, applying to quite a few important polyelectrolytes, like
DNA, actin filaments or microtubules. We treat the solvent in
dielectric approximation and explicitely describe only the small ions.
Within Poisson-Boltzmann (PB) theory \cite{gouy10a} and on
the level of a cell model the cylindrical geometry can be treated
exactly in the salt-free case
\cite{alfrey51a,katchalsky71a,wennerstroem82a,lebret84a,lebret84b,deserno00a},
providing for instance new insights into the phenomenon of the Manning
condensation \cite{manning69a,oosawa71a}.

The paper is organized as follows: First we review some essential
features of the PB mean-field solution of the cell model. Then we
discuss the applicability of PB theory to the ion distribution
functions and show under which circumstances PB theory fails
quantitatively (underestimated condensation) and qualitatively
(overcharging, charge oscillations and attractive interactions).  Next
we present measurements of the osmotic pressure for the nematic phase
of hexagonally packed polyelectrolytes as a function of density and
compare it to PB predictions and the Manning limiting law. In the next
section we study the particular case of poly(para-phenylene) by means
of PB theory, including a correlation correction of the basis of a
recently proposed Debye-H\"uckel-Hole-cavity theory (DHHC)
\cite{barbosa00a}, and simulational results. The results are compared
to recent experimental data on this system
\cite{guilleaume00a,blaul00a,deserno01a}. We find that correlation
effects enhance condensation and lower the osmotic pressure, yet are
not fully able to explain the discrepancy to the experimental data.
At the end we try to shed light onto the role of specific ionic
correlations. Two-dimensional correlations on the surface of the rods
are found to be present, but weakly developed. No hexatic order of the
ions is observed.


\section{Simulation method and model system}\label{sec:model_system}
%
%
\subsection{The Langevin thermostat}\label{sec:genalpre}
We utilize molecular dynamics (MD) simulations using a Langevin
thermostat \cite{grest86a} to study the equilibrium properties of our
model system within the canonical ensemble. Technically this is
achieved by integrating the stochastic differential equation
\begin{equation}
  m_i \ddot{\VECr}_i \; = \; -\nabla U(\{\VECr_i\}) - \Gamma\dot{\VECr}_i +
  \VECxi_i(t)
  \label{eq:Langevin_equation}
\end{equation}
on the computer, where $m_i$ are the masses of particles with
coordinates $\VECr_i$ subject to a potential energy function $U$, a
friction $\Gamma$ linear in the velocity and a stochastic white noise
$\VECxi_i(t)$. The latter two can be thought of as imitating the
presence of a surrounding viscous medium responsible for a drag force
and random collisions, respectively. Since both effects have the same
origin, they are related to each other by a simple version of the
fluctuation-dissipation-theorem. This can be exploited to choose their
strength such as to converge towards the canonical state:
\begin{equation}
  \langle \VECxi_i(t) \rangle = 0 
  \qquad \text{and} \qquad
  \langle \VECxi_i(t)\cdot\VECxi_j(t') \rangle = 
  6 \, k_\rB T \, \Gamma \delta_{ij} \delta(t-t'),
  \label{eq:while_noise_moments}
\end{equation}
where $k_\rB T$ is the thermal energy.

We use a standard Verlet integrator\cite{allen87a} to
integrate (\ref{eq:Langevin_equation}). Time step $\delta t$ and
friction coefficient $\Gamma$ were set to 0.01 and 1.0 in
Lennard-Jones units (see below).
%
\subsection{Interaction potentials}
We use the purely repulsive Lennard-Jones (LJ) potential to give the
counterions an excluded volume:
\begin{equation}
  V_\rLJ(r) \; = \; \left\{
    \begin{array}{r@{\quad:\quad}l}
      4\epsilon\left[\left(\D\frac{\sigma}{r}\right)^{12} \!-\;
        \left(\D\frac{\sigma}{r}\right)^6 + \D\frac{1}{4}\right] & 0 < r \le
      r_\rcut\equiv 2^{1/6}\sigma \\[2ex]
      0 \hspace*{12ex} & r_\rcut < r.
    \end{array}
  \right.
  \label{eq:LJ_potential}
\end{equation}
The advantage of including the $-r^{-6}$ contribution instead of
merely using the purely repulsive $r^{-12}$ is that
Eqn.~(\ref{eq:LJ_potential}) is exactly zero beyond $r_\rcut$ and
merges smoothly to this value at $r_\rcut$, allowing a larger time
step.

The Coulomb potential of a charge $Q$ is written as
\begin{equation}
  \beta e_0 V_\rC(r) \; = \; (Q/e_0)\,\frac{\ell_\rB}{r},
\end{equation}
where $\ell_\rB=\beta e_0^2 / 4\pi\varepsilon$ is the Bjerrum length, the
distance at which two unit charges have an interaction energy equal to $1/
\beta := k_\rB T$, $e_0$ is the positive elementary charge, and $\varepsilon$
is the product of the dielectric constants of vacuum, $\varepsilon_0$,
and the medium, $\varepsilon_\rr$, respectively.  The Lennard-Jones
unit $\sigma$ is always used as unit of length and $\epsilon$ is
always set to the thermal energy $k_\rB T$.  Temperature is
implemented via the Bjerrum length.  Mass is irrelevant and set equal
to one -- it would only be needed to translate the Lennard-Jones time
$\tau_\rLJ=\sigma\sqrt{m/\epsilon}$ into ``real'' time.  We intend to
model an aqueous environment at room temperature, which implies
$\ell_\rB\approx 7.14\,\text{\AA}$.

Within the periodic boundary conditions employed during the
simulations, the presence of such long-range interactions poses both
mathematical and technical difficulties. We use the most efficient FFT
accelerated Ewald sum, the P$^3$M code, which scales almost linearly with
the number of charges \cite{deserno98a,deserno98b}.
%
%
\subsection{Generating a cell-geometry}
Compared to the spherical cell model, the cylindrical one presents one
additional but crucial complication: the charged rod is infinitely long.
Several methods have been proposed in the literature to handle this problem
\cite{guldbrand89a,nilsson91a,jensen97a,lyubartsev98a}. They essentially
all use as a unit cell a hexagonal prism with a certain height.  This
approximates the cylindrical cell by a space-filling object. In the
present work we take a cube of side length $L_\rb$ and place a rod
along the main diagonal. Upon periodically replicating this system the
diagonal rod becomes infinitely long and an infinite triangular array
of such replicated rods emerges.  The resulting Wigner-Seitz cell of
this lattice is a regular hexagon, which can alternatively be viewed
as the unit cell. This is illustrated in Fig.~\ref{pic:cellpicture}.
Observe that the symmetry of the replicated system is still cubic. The
radius $R$ of a circle with the same area $A$ is then given by
$R=L_\rb/\sqrt{\pi\sqrt{3}}$. This value is most appropriately used
for comparing results between the hexagonal and the cylindrical cell
model.

If the line charge density of the rod is $\lambda$, electroneutrality
requires the number of $v$-valent counterions to be $N =
\sqrt{3}L_\rb\lambda/ve_0$.  Hence the average counterion density is
given by $n = N/L^3=\sqrt{3}\lambda/ve_0L^2$ and is thus inversely
proportional to $L^2$ instead of $L^3$. The number of counterions can
therefore be written as $N=(\sqrt{3}\lambda/ve_0)^{3/2}n^{-1/2}$,
which implies that a smaller density requires more particles.  While
this makes the simulation of dilute systems rather expensive, it gives
at the same time rather small dense systems. The latter problem can be
circumvented by combining blocks of $2 \times 2 \times 2$, $3 \times 3
\times 3$ or even more elementary cubes to a big cube and using the
latter as the unit box for the periodic boundary conditions.
  
The ratio between $\ell_\rD = (4\pi\ell_\rB v^2 n)^{-1/2}$ (the
average Debye length of the counterions) and the rod separation $d_\rrod$
can be written as
\begin{equation} 
\frac{\ell_\rD}{d_\rrod} =  
(8\pi\xi v / \sqrt{3})^{-1/2} \; \approx \; 0.2625/\sqrt{\xi v},
\end{equation} 
where the definition of the Manning parameter $\xi :=\lambda \ell_\rB
/ e_0$ was used, see also Sec.~\ref{Sec:PB}.  Obviously, for strongly
charged rods the Debye length is smaller than the separation of the
two rods. Even for $\xi v=1$ it is only half as large as the distance
between rod axis and Wigner-Seitz boundary, and the neighboring cells
effectively decouple. This statement is independent of density.
%
\section{Poisson-Boltzmann Essentials}\label{Sec:PB}
%
%
\subsection{The Poisson-Boltzmann Solution}
%
%
At this stage we want to briefly recall the necessary knowledge about
the PB solution of the cylindrical cell model
\cite{alfrey51a,wennerstroem82a,lebret84a,lebret84b,deserno00a}.
Consider an infinitely long cylinder of radius $r_0$ and line charge
density $\lambda>0$, coaxially enclosed in a cylindrical cell of
radius $R$.  Global charge neutrality of the system is ensured by
adding an appropriate amount of oppositely charged $v$-valent
counterions. In the following only the case of no extra salt will be
discussed.

Within PB theory the individual counterions are replaced by a
cylindrical counterion density $n(r)$, where $r$ denotes the radial
distance from the cylinder axis.  Defining the reduced electrostatic
potential $y$ and and the screening constant $\kappa>0$ as
\begin{equation}
  y(r) \; = \; \beta e_0 v \psi(r)
  \qquad\text{and}\qquad
  \kappa^2 \; = \; 4\pi\ell_\rB v^2 n(R),
\end{equation}
the PB equation can be written as
\begin{equation}\label{Poisson_Boltzmann}
  y'' + \frac{y'}{r} \; = \; \kappa^2 \exp(y).
\end{equation}
It will be useful to define the dimensionless \emph{Manning parameter}
$\xi = \lambda \ell_\rB / e_0$ which counts the number of unit
charges on the rod per Bjerrum length. In the following the main focus
will be on the strongly charged case characterized by $\xi v>1$.
Fixing the two boundary conditions for the electric field,
\begin{equation}\label{boundary_E_field}
  y'(r_0) = - 2 \xi v / r_0
  \qquad\text{and}\qquad
  y'(R) = 0,
\end{equation}
the correctly normalized solution to
Eqns.~(\ref{Poisson_Boltzmann},\ref{boundary_E_field}) can be written as
\cite{alfrey51a,lebret84b,deserno00a}
\begin{equation}\label{PB_potential}
  y(r) = -2\,\ln\left\{\frac{r}{R}\,\sqrt{1+\gamma^{-2}}\,
    \cos\Big(\gamma\,\ln\frac{r}{R_\rM}\Big)\right\}.
\end{equation}
Insertion of the general solution from Eqn.~(\ref{PB_potential}) into
the boundary conditions from Eqn.~(\ref{boundary_E_field}) yields two
coupled transcendental equations for the two integration constants
$\gamma$ and $R_\rM$:
\begin{equation}\label{r0_R_RM}
\gamma\,\ln\frac{r_0}{R_\rM} = \arctan\frac{1-\xi v}{\gamma} \qquad\text{and}\qquad
\gamma\,\ln\frac{R}{R_\rM} = \arctan\frac{1}{\gamma}.
\end{equation}
Subtracting the left part from the right part in Eqn.~(\ref{r0_R_RM}) eliminates
$R_\rM$ and provides a single equation
\begin{equation}\label{gamma_equation}
  \gamma\,\ln\frac{R}{r_0} = 
  \arctan\frac{1}{\gamma} + \arctan\frac{\xi v-1}{\gamma},
\end{equation}
from which $\gamma$ can be obtained numerically. The second integration
constant $R_\rM$, which will be referred to as the Manning radius, can be
obtained from either one of the Eqns. in (\ref{r0_R_RM}) as soon as $\gamma$
is known. Note finally that $\kappa$ and $\gamma$ are connected via
\begin{equation}\label{kappa_gamma}
\kappa^2 R^2 = 2 \, (1+\gamma^2).
\end{equation}

Since the $\xi$ and $v$ only enter in the combination $\xi v$,
changing valence or electrostatic interaction strength is equivalent
on Poisson-Boltzmann level. In particular, at given cell geometry
$\{r_0,R\}$ the Manning radius $R_\rM$ is a unique function of $\xi
v$.  Eqn. (\ref{gamma_equation}) implies the sequence of inequalities
$\ln(R/r_0) \le \pi/\gamma \le \ln(R/r_0) + \xi v/(\xi v-1)$. The
resulting asymptotic for $\gamma$ in the dilute limit $R \rightarrow
\infty$ gives rise to what is often called the ``Manning limiting
laws''.

The integrated probability distribution of finding a mobile ion at distance
$r$ within a cylinder of radius $r\in[r_0;R]$ can be determined analytically
by integrating the charge density:
\begin{equation}
  P(r) \; = \; \frac{1}{\lambda} \int_{r_0}^r \infd r' \, 2\pi r' \, 
  e_0 \, v \, n(r') \; = \; \Big(1 - \frac{1}{\xi v}\Big) +
  \frac{\gamma}{\xi v}\tan\Big(\gamma\,\ln\frac{r}{R_\rM}\Big).
\label{probdistri}
\end{equation}
This is the fraction of counterions found within a cylinder of radius
$r$. $P(r)$ will be referred to as counterion distribution
function. At $r=R_\rM$ the last term in $P$ vanishes, giving the
Manning fraction $f_\rM := 1-1/\xi v$ of ions within $R_\rM$. These
are the ions which can not be diluted away, if the cylinder radius is
sent to infinity.  This phenomenon is referred to as {\em Manning
condensation} \cite{manning69a,oosawa71a}.
%
%
\subsection{How to quantify counterion condensation}\label{ssec:infpoint}
If the counterion distribution function $P$ is known, the condensed
counterion fraction can be characterized in the following
``geometric'' way: Eq.~(\ref{probdistri}) shows that $P$ viewed as a
function of $\ln(r)$ is merely a shifted tangent-function with its
center of symmetry at $(\ln(R_\rM);f_\rM)$.  Since, however,
$\tan''(0) = 0$, Manning radius and Manning fraction can be found by
plotting $P$ as a function of $\ln(r)$ and localizing the point of
inflection.

This property of $P$, derived within the framework of PB theory, can
in turn be used to {\em define} the condensed fraction
\cite{belloni84a,deserno00a}.  It provides a suitable way to quantify
counterion condensation beyond the scope of PB theory, and it is exact
in the salt-free PB limit.  From here on this method will be referred
to as the {\em inflection point criterion}.  It has the advantages of
{\em (i)} not fixing by definition the amount of condensed counterions
($f_\rM$ and $R_\rM$ can be determined independently of each other),
{\em (ii)} reproducing the salt-free PB limit, and {\em (iii)}
quantifying the breakdown of the coexistence of condensed and
uncondensed counterions in the high salt limit.
%
%
\section{Generic ion distribution functions}\label{sec:generic_distrifunc}
On the plain PB level the radius $r_0$ of the rod is not a completely
independent variable, since it enters only in the combination $R/r_0$,
see Eqn. (\ref{gamma_equation}). The rod in our simulation is built up
from a sequence of small Lennard-Jones particles lined up along the
main diagonal of the simulation cell.  The distance of closest
approach of two Lennard-Jones particles can then be identified with
the radius $r_0$ of the rod and yields an effective rod radius of
$\sigma$.

Upon leaving the PB level, the ratio between counterion diameter
$\sigma$ and rod radius $r_0$ becomes relevant, which has for instance
been investigated in Ref.~\cite{jensen97a,allahyarov00a}. Here we do
not intend to perform a systematic investigation of such effects, but
instead present later in Sec.~\ref{sec:DNAppp} results for systems
mapped to physically relevant parameters, DNA and two kinds of
poly(p-phenylenes), in which the rod has a considerably larger
diameter than the counterions.
%
%
\subsection{Density dependence within monovalent systems}\label{ssec:nscan_mono}
At fixed rod radius of $r_0$ the relevant variable $R/r_0$ is changed
by varying the cell size $R$ and therefore the density.  Here we
present results for such a density scan for systems with
$\ell_\rB/\sigma \in \{2,3\}$. The monovalent and positively charged
particles forming the rod were placed along the main diagonal with a
center-center distance of $1.04245\,\sigma$, giving a line charge
density of $\lambda=0.959279\,e_0/\sigma$. In connection with the two
presented values for the Bjerrum length this yields Manning parameters
of $\xi = 1.919$, $2.878$ and corresponding Manning fractions of
$f_\rM = 1-1/\xi = 0.4788$, $0.6525$ in the monovalent case. The cell
radius $R$ has been varied between $2.06 \sigma$ and $124 \sigma$,
which for monovalent counterions corresponds to average ion densities
of $7.2 \times 10^{-2} \sigma^{-3}$ and $2.0 \times 10^{-5}
\sigma^{-3}$, respectively.  The inflection point criterion from
Sec.~\ref{ssec:infpoint} has been used to determine the radial
extension of the condensed layer, $R_\rM^\star$, and the fraction of
ions within it, $f_\rM^\star=P(R_\rM^\star)$, where the star denotes
the measured quantities. A graphical illustration of the distribution
functions is given in Fig.~\ref{pic:mono_dens_scan}.

The important things to observe are the following. The measured
condensed fraction $f_\rM^\star$ is always larger than the fraction
predicted by the PB theory. However, the fraction from the simulation
decreases monotonically with decreasing density towards the Manning
limit $f_\rM$. Such a deviation is to be expected, since PB is
essentially a low density theory \cite{deserno00b,deserno00c}.

In contrast to the clear tendency of the measured condensed fraction
to decrease upon dilution, the behavior of the condensation radius
$R_\rM^\star$ appears to be more complicated. There does not seem to
exist a simple monotonic convergence of $R_\rM^\star$ towards
$R_\rM$. Rather, for high densities the measured condensation distance
is larger than the Manning radius, while for the investigated low
densities it is smaller. Unfortunately, a clear-cut statement is
difficult since the localization of the point of inflection in $P$ as
a function of $\ln(r)$ is only possible with an error estimated to be
of the order of $1\%$.

For Bjerrum length $\ell_\rB=1\,\sigma$, corresponding to $\xi=0.9593<1$, the
radial distribution functions are compared to the PB predictions in
Fig.~\ref{pic:mono_dens_scan_Bj1}.  It can be seen that the measured and PB
predicted distributions almost coincide for all cell sizes under
investigation. While this is to be expected for low densities, the remarkable
agreement at high densities is somewhat surprising.  Apparently, PB theory is
a fairly good description of weakly charged systems, i.e., ones which are
below the Manning threshold.
%
%
\subsection{Fitting to a generalized Poisson-Boltzmann distribution}\label{ssec:fitPB}
Since the measured Manning radius $R_\rM^\star$ can be smaller than
the PB prediction, but $f_\rM^\star>f_\rM$ and $R_\rM^\star<R_\rM$ is
incompatible with PB theory, it is impossible to describe such a
measured $P(r)$ with a PB distribution function in which a somewhat
enlarged effective Manning parameter $\xi^\star$ is used.
Nevertheless, it would be desirable to use at least some of the
knowledge about the ion distribution function from PB theory for
evaluating relevant observables in a simulation or in a real
experiment.  To the lowest order, e.g., within a two state model, the
increase of counterion condensation can be ascribed to such an
effective $\xi^\star v = 1/(1-f_\rM^\star)$. To break the monotonic
connection between $f_\rM$ and $R_\rM$ one assumes that the functional
form of $P$ is given by the PB form in Eqn.~(\ref{probdistri}), but
neglects the relation in Eqn.~(\ref{gamma_equation}). Using
Eqn.~(\ref{r0_R_RM}{\em i}), this suggests fitting the measured
distribution in a region around the inflection point to the form
\begin{equation}
  P_\rfit(r) \; = \;
  1-\frac{1}{\xi^\star v} + \frac{\gamma^\star}{\xi^\star v}
  \tan\bigg(\gamma^\star\,\ln\frac{r}{r_0^\star} + 
  \arctan\frac{1-\xi^\star v}{\gamma^\star}\bigg)
  \label{eq:P_fit}
\end{equation}
with the three fit parameters $\xi^\star$, $\gamma^\star$ and
$r_0^\star$.\footnote{It might seem strange to regard the rod radius as a free
  parameter, but there are two reasons for this. First, this does not force
  the fit to coincide with the PB form outside the chosen fitting-region.
  Second, not even in a real experiment is the radius of a macroion always
  known in advance. Rather, it is often necessary to obtain this information
  within the same measurement \cite{guilleaume00a}.} The condensed fraction is
then given by $f_\rM^\star=1-1/\xi^\star v$ and the condensed radius by
$R_\rM^\star=r_0^\star\exp\{\arctan[(\xi^\star v-1)/\gamma^\star]/\gamma^\star\}$.
Fig.~\ref{pic:mono_PB_fit} illustrates this procedure for one system. While
this approach might seem to be very powerful, it suffers from the drawback
that the actual numbers depend on the chosen fitting region.  Nevertheless, it
provides an independent way of quantifying condensation and can even be
applied to very noisy data, for which the localization of an inflection point
in $P$ as a function of $\ln(r)$ is otherwise virtually impossible.
%
%
\subsection{Multivalent ions}
The universal dependence of the ion distribution on the product $\xi
v$ is an artifact of Poisson Boltzmann theory.
Fig.~\ref{pic:mono_bi_tri} shows examples of systems with different
Manning parameters but the same value of $\xi v$. Not only is the
condensation enhanced as compared to PB theory, but the enhancement is
stronger for the case involving multivalent ions. For instance, in the
dense trivalent system the condensation is enhanced by 30\% with
respect to the PB prediction.

The reason is that at given charge density multivalent systems have a lower
number density and thus fewer particles. This lowers any kind of excluded
volume interactions, and also makes the entropic contribution less pronounced.
Moreover, the relevant variable for describing the strength of correlation
effects is $\ell_\rB v^2$ \cite{barbosa00a}. Hence, an increase in
valence is more important than an increase in Bjerrum length.

Within PB theory the expression for the contact density $n(r_0)$ in
the limit of infinite dilution is given by
\begin{equation}
  n(r_0) \; \stackrel{R\rightarrow\infty}{=} \;
   2\pi\ell_\rB\tilde\varsigma^2\;\bigg(1-\frac{1}{\xi v}\bigg)^2,
  \label{nr0v_limit}
\end{equation}
where $e_0\tilde\varsigma$ is the surface charge density of the rod.
For high rod charge this expression becomes independent of
valence. Hence, replacing monovalent counterions by multivalent ones
reduces their total \emph{number} in the cell, but for a highly
charged rod not their \emph{density} at the rod surface. This result
is a consequence of the contact value theorem and thus holds beyond
the Poisson-Boltzmann approximation\cite{wennerstroem82a}.
%
%
\subsection{Addition of salt}\label{ssec:add_salt}

The influence of salt on the distribution functions has been discussed
within the PB theory in detail in Ref. \cite{deserno00a}. The general
finding was that a low salt content leaves the picture of Manning
condensation qualitatively unchanged, while at increasing salt
concentration a crossover between Manning condensation and simple
salt-screening occurs.

If one starts adding $N$ salt molecules to the salt free solution one
finds that the inflection point gets shifted to smaller values of $r$,
hence the layer of condensed counterions contracts.  The second
finding is that the {\em amount} of condensation is only marginally
increased.  From a certain $N$ on two more inflection points appear
close to the cell boundary. This happens typically for a corresponding
Debye length being of the order of the cell size itself, indicating
the appearance of a characteristic, salt induced, change in the
convexity of $P$ (as a function of $\ln r$).  Upon a further increase
in $N$ one of the two new inflection points shifts towards smaller
values of $r$, finally fusing with the Manning inflection point
and ``annihilating'' it. Roughly speaking, the inflection points
vanish if the Debye length characterizing the salt content becomes
smaller than the radius of the condensed layer. In this case it is no
longer meaningful to distinguish between condensed and uncondensed
counterions.  Indeed, for a very high salt content, where the Debye
length is much smaller than the radius of the rod, the solution of the
PB equation would be the one of a charged plane and one may consider
all excess counterions being condensed -- no matter what the charge
density of the rod is.

For three systems with numbers of salt molecules $N\in\{0,104,3070\}$
a simulation has been performed and compared to the PB prediction in
Fig.~\ref{pic:SW575859}.  As in the salt-free case the computer
simulations show a more pronounced condensation effect towards the
rod.  Nevertheless, the shape of the distribution functions remains
qualitatively the same.  Note in particular that the appearance and
disappearance of two points of inflection at $N=104$ and $N=3070$
respectively, which leads to extremely small curvatures in the PB
distribution functions, also leads to very straight regions in the
measured distribution functions. The crossover from Manning
condensation to screening, as described within the PB theory, can be
expected to be essentially correct.  It should not be overlooked,
however, that the addition of salt ions will affect the distribution
function much more dramatically if their valence is larger than the
valence of the counterions.  In this case, the salt counterions will
accumulate in the vicinity of the rod at the expense of the lower
valent ``real'' counterions as has previously been demonstrated in a
system with a valence mixture of counterions
\cite{deserno00d}.

The PB approach fails to describe the physical situation if one or
more of the following conditions apply: {\em (i)\/} the electrostatic
interactions are strong, {\em (ii)\/} the counterions are multivalent
or {\em (iii)\/} the density is high. Results of a simulation under
such conditions can be seen in Figure~\ref{pic:challenge}.  Here, a
system with box length $L_\rb/\sigma=36.112$, i.e., 60 monovalent
counterions and a cell size of $R/\sigma =15.481$, and
$\ell_\rB/\sigma=4.1698$, i.e., $\xi=4$, has been investigated after
adding 1000 molecules of a 2:2 salt. Since this gives almost 17 times
as much salt as counterions and a salt Debye length of
$\ell_\rD/\sigma=0.33\ll R/\sigma$, this can essentially be viewed as
a charged rod in a bulk 2:2 electrolyte. The most characteristic
feature of the charge distribution function is that it overshoots
unity, showing a charge reversal of the rod at distances around $r
\approx 1.5\,\sigma$, while the simple PB prediction is clearly
qualitatively off.  This phenomenon is usually referred to as
overcharging and has been predicted for the rod geometry first
from hypernetted chain calculations \cite{gonzalestovar85a} and later
by a modified PB approach \cite{das95a,das97a}, and has also been observed in
planar geometries\cite{greberg98a}.  Since $P(R)=1$ for
the reason of global electroneutrality, the overshooting above 1 at
small distances implies the existence of a range of $r$-values at
which the mobile ion system is locally positively charged, i.e., with
the same charge as the rod, such that $P(r)$ can eventually decay to
1. This is seen in the right frame of Fig.~\ref{pic:challenge}, which
shows that $n_{+2}(r)>n_{-2}(r)$ at $r \approx 2\,\sigma$. Since
$P(1.5\,\sigma)\approx 1.45$, the rod and its innermost layer of
condensed ions could be viewed as an effective rod of radius
$1.5\,\sigma$ which is negatively charged with Manning parameter
$\xi=1.8$.  Since this value is again larger than 1, it entails ion
condensation, but this time of positive ions.  In fact, it even leads
to a second overcharging, as can clearly be seen in
Fig.~\ref{pic:challenge}, where $P(r)$ -- in decaying from 1.45 --
overshoots the value of 1 again.  Overcharging can thus give rise to
layering. In the presented example no less than three layers can
clearly be made out. These local charge oscillations also reflect
themselves in oscillations of the electrostatic potential, as
demonstrated in the inset in the right frame of
Fig.~\ref{pic:challenge}.  Notice that these oscillating potentials
will also have pronounced effects on the interaction between such
rigid polyelectrolytes. HNC/MSA theory can qualitatively and
quantitatively describe such effects to a very good degree, as has
been demonstrated theoretically a long time ago
\cite{gonzalestovar85a}.  A recent comparison with simulations can be
found in Ref.~\cite{deserno01b}.


\section{Pressure}\label{sec:pressure}


\subsection{Pressure measurements within the generic
  model}\label{ssec:pressgeneric}

Using the results from the Appendix \ref{ssec:defpress} we now turn
towards pressure measurements for the generic systems described
above. Because of the periodic boundary conditions, we are actually
looking at the pressure of a nematic, hexagonally packed solution of
charged rods. We specifically compute the dimensionless osmotic
coefficient $\oscoeff$, which is simply the pressure normalized by its
ideal gas contribution
\begin{equation}
  \oscoeff \; = \; \frac{p}{p^\rig} \; = \; \frac{p}{n k_\rB T}.
\end{equation}
For an {\em isolated\/} cell the pressure is given by the particle
density at the outer cell boundary. This is a rigorous statement, true
for the spherical, cylindrical and planar cell model
\cite{wennerstroem82a}. It is a merit of the PB equation that it
retains the validity of this exact relation. For the extended density
functional theories this no longer holds and additional terms appear
\cite{deserno00b}.

Within Poisson-Boltzmann theory the osmotic coefficient as determined
from the boundary density is
\begin{equation}
  \oscoeff \; = \; \frac{1+\gamma^2}{2\xi v} \;
  \stackrel{R\rightarrow\infty}{=} \; \frac{1}{2\xi v},
  \label{eq:oscoeff}
\end{equation}
where $\gamma$ is the density dependent integration constant from
Eqn.~(\ref{gamma_equation}),

Fig.~\ref{pic:press_Bj123} illustrates these measurements for
monovalent counterions and three values of $\ell_\rB/\sigma = 1,2,3$.
Several things may be noted: The osmotic coefficient from the
simulations is always smaller than the PB prediction, but for low
density both values converge. This also illustrates that the Manning
limiting law from the r.h.s of Eqn.~(\ref{eq:oscoeff}) becomes
asymptotically correct for dilute systems.  Upon increasing the
density, the osmotic coefficient rises weaker than the PB
prediction. This is more pronounced for systems with higher Bjerrum
length, and consequently, higher Manning parameter, and is due to
enhanced counterion condensation which has been observed in
Sec~\ref{sec:generic_distrifunc}.  Notice that this has a very
remarkable side-effect.  Over a considerable range of densities the
measured osmotic coefficient is much closer to the limiting law than
to the actual PB prediction.  This makes the Manning limit look much
more accurate than it really ought to be. However, the surprising
effect should not be over-interpreted, since the underlying reason is
nothing but a fortunate cancellation of two contributions of
approximately the same size which are not contained in the limiting
laws.

Also for the pressure it is interesting to investigate complementary
systems in which the values of Bjerrum length $\ell_\rB/\sigma$ and
valence $v$ have been interchanged to keep the product $\xi v$ fixed.
However, at given density the cell radius depends on the valence, so
$\oscoeff(n)$ does no longer universally depend on this product. Since
in the limit of high density $\oscoeff$ is easily seen to increase
with density, while at low density it decreases, the functions
$\oscoeff(n)$ for different valences intersect at some point.

Fig.~\ref{pic:press_v123} summarizes the results of measurements on
the multivalent systems with $v = 1,2,3$, which yield the same values
of $\xi v$ as the monovalent ones investigated before.  The most
striking feature is the appearance of a negative osmotic coefficient
in a certain density region of the trivalent case.  If the constraint
of a fixed rod-separation were to be replaced, the system would phase
separate, hence, attractive interactions must be present between the
rods.  Similar observations have been reported in
Refs.~\cite{nilsson91a,jensen97a,lyubartsev98a}. In order to trace
back the origin of those attractions, Fig.~\ref{pic:press_v23_parts}
displays the osmotic coefficient in the divalent and trivalent cases,
split into two contributions: {\em (i)\/} the non-electrostatic part
comprising ideal gas contribution and the short-range virial and {\em
(ii)\/} the electrostatic part, which for visual convenience is
plotted with a reversed sign. Hence, the difference between those two
curves gives the curves in Fig.~\ref{pic:press_v123}.  We observe that
the electrostatic part leads to a monotonically increasing attraction
with increasing density. It is almost twice as strong in the trivalent
case.  The very strong increase of the osmotic coefficient at large
densities is due to the virial, i.e., due to repulsive ion-rod and
ion-ion interactions.  The measured negative pressure in the trivalent
case is the result of a ``sudden'' drop in the virial
contribution. Nothing particular can be observed in the electrostatic
part.  We suggest the explanation that in an intermediate range of
densities the presence of surrounding rods lifts condensed ions from
the surface they are pushing on, thereby reducing the short range
excluded volume contribution. In too dilute systems this effect is not
operative, while in too dense systems the effect of the other rods is
actually to push ions more closely to the surface. Hence, there is a
density range in which the virial is reduced by the presence of other
rods.

Contrary to the simulations, the osmotic coefficient from the PB
theory is always positive. This is the consequence of the rigorous
proof that such attractive interactions are absent on the PB level
\cite{neu99a}. An extension of this statement to ions of
finite size and to a wider class of boundary conditions and density
functionals can be found in Ref.~\cite{trizac00a}.

Finally it should be noted that the above measurements can not be used
to infer that attractive forces between charged rods require the
counterions to be at least trivalent.  The reason is twofold: First,
at given valence one can vary Bjerrum length and line charge
density. Increasing the Manning parameter will lead to negative
pressure in the divalent system.  Second, keeping all interaction
potentials fixed, the radius $r_0$ of the charged rod is a relevant
observable, as has been demonstrated in Ref.~\cite{jensen97a}.  Hence,
a general statement about presence or absence of attractive
interactions is difficult, since a five-dimensional parameter space is
involved: $\{\lambda, \ell_\rB, v, r_0, n\}$.


\section{Examples: poly(p-phenylene)}\label{sec:DNAppp}

This section discusses simulations in which the parameters are
explicitly mapped to an experimental system. This affects the values
of ion diameter $\sigma$, rod radius $r_0$, Bjerrum length $\ell_\rB$
and line charge density $\lambda$. In order to have $r_0$ different
from $\sigma$, we use as the ion-rod-potential a LJ like expression,
in which $r$ is replaced by $r-r_\rs$ and the shift $r_\rs$ is related
to the desired rod radius by $r_0=r_\rs+\sigma/2$.

Since these systems have been investigated either in the presence of
salt or at rather high Manning parameter, the phenomena happening in
the vicinity of the rod are always well decoupled from the cell
boundary. Therefore, an even simpler geometry has been chosen: One rod
placed parallel to one of the edges of a cubic simulation cell.


\subsection{Distribution functions and osmotic coefficient
for salt free poly(p-phenylene) solutions}

\begin{figure}[h]
\begin{center}   \includegraphics{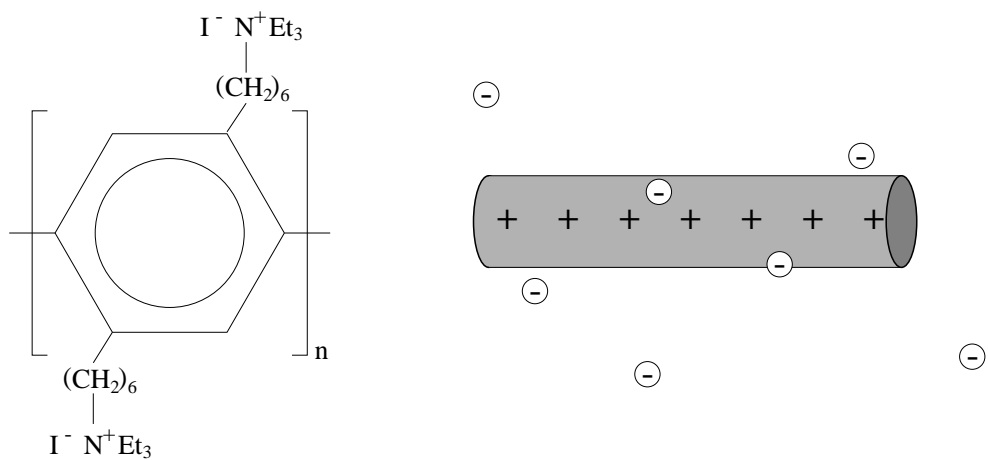} \end{center}
    \caption{Example of a stiff polyelectrolyte.
    Constitution formula for poly(para-phenylene) with iodine counterions
    (left) and a physicist's picture (right).}\label{Fig2}
\end{figure}

\begin{figure}
\begin{center}  
\includegraphics{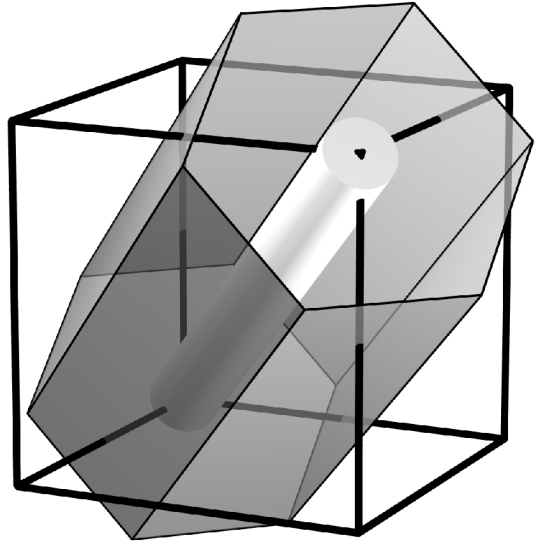}  
\end{center}
 
    \caption{Realization of the cell model. A rod placed along the main
      diagonal of a cube yields an infinite triangular array of infinitely
      long rods upon periodic replication of the original cube. The
      Wigner-Seitz cell of this lattice is a regular hexagon enclosing the
      rod. This therefore provides a way of obtaining a hexagonal cell without
      abandoning a cubic geometry.}\label{pic:cellpicture}

\end{figure}

The Manning limiting laws are claimed to apply in the limit of low
density.  However, an experimental verification of this effect using
DNA as rod-like polyelectrolytes is difficult, since the double helix
starts to unwinds at low ionic strength.  It would therefore be
desirable to have a stiff model polyelectrolyte which does not suffer
from this problem. One such system belonging to the class of
poly(para-phenylenes) (ppp) has recently been investigated in
Refs.~\cite{guilleaume00a,blaul00a}. The constitution formula is given
in Fig.~\ref{Fig2}. The fully aromatic backbone exhibits an excellent
chemical stability and has a persistence length of approximately
$20\,\text{nm}$. The degree of polymerization used in the
abovementioned studies was between 20 and 40, so that the contour
length equals approximately one persistence length at most. The
dominant contribution to the signal in small angle X-ray scattering
experiments stems from the iodine ions, since the excess electron
density of the backbone is very low. Table~\ref{tab:ppp} lists four
systems which have been simulated based on a mapping to ppp.

In a first step the simulated ion distribution functions shall be
compared with the PB prediction as well as with the
Debye-H\"uckel-hole-cavity theory from Ref.~\cite{barbosa00a}, which
tries to incorporate correlations missing in the PB
approach. Figure~\ref{pic:birgit_Pr} shows the corresponding
distribution functions for the systems from Table~\ref{tab:ppp}.
Qualitatively the PB prediction is already a good description of those
systems, if one graciously disregards the inevitable problems at the
cell boundary. However, the presence of correlations shifts the
distribution functions up in all four cases. Since this shift is
rather small, i.e., the correlations are weak, the
Debye-H\"uckel-hole-cavity theory is in fact an excellent description
of those systems. Observe, for instance, that in the weakly charged
dilute case its prediction can no longer be distinguished from the
simulated curve on the chosen scale.

With the help of the condensation criterion from
Sec.~\ref{ssec:infpoint} the extent of the correlation-enhanced
condensation can be further quantified.  The Manning fraction from the
molecular dynamics simulation increases only by a fairly small
amount. It is at most 4\% larger than the PB value. This translates to
an effective Manning parameter $\xi_\eff=1/(1-f_\rM)$ being
$5$--$10$\% larger than the bare one. This increase is very accurately
captured by the Debye-H\"uckel-hole-cavity theory.  Its prediction for
$f_\rM$ is at most 1\% smaller than the value obtained in the
simulation. Interestingly, it is even independent of density. This,
however, is not a feature to be generally expected and should
therefore not be over-interpreted.

In a second step the osmotic coefficients of the four model systems have been
computed. Two distinct approaches have been used for analyzing the simulation
results: The first computes the pressure from the stress tensor, as described
in detail in Appendix~\ref{ssec:defpress}.  The second approach exploits the
connection from Eqn.~(\ref{eq:oscoeff}) between osmotic coefficient, Manning
parameter and the integration constant $\gamma$ of the PB equation. The idea
is to use the fitting procedure described in Sec.~\ref{ssec:fitPB} and from
that obtain values $\xi^\star$ and $\gamma^\star$, leading to the osmotic
coefficient $(1+{\gamma^\star}^2)/2\xi^\star$. The most straightforward
approach of measuring the boundary density is less recommendable, since in the
simulation the boundary has a quadratic and not a circular cross section,
i.e., the merely approximate representation of the cell model becomes most
visible and questionable here. In contrast to that, the two other approaches
``look'' at regions of the cell, which are away from the boundary. For
comparison, also the predictions from the Manning limiting law and the full PB
expression are calculated. For the two dilute systems there also exist
measurements in Ref.~\cite{blaul00a} of $\oscoeff$ via osmometry, while for
the dense systems this was unfortunately impossible due to counter diffusion
problems.  Table~\ref{tab:ppp_oscoeff} collects the coefficients for
comparison.

The first thing which should be noted is that PB theory gives again a
surprisingly good description of the measured coefficients, including
the experimentally determined one. A closer look reveals that -- as
expected -- it overestimates $\oscoeff$, and the deviations are larger
for the dense systems.  Still, PB theory is off by only 7\% or 18\%
for the dilute or dense systems, respectively, thereby confirming the
observation already made when looking at the distribution functions. A
further indication of this point is that the Manning limit $1/2\xi$
appears to be a lower boundary, which is also in accord with the
pressure measurements on the generic systems with monovalent ions from
Sec.~\ref{ssec:pressgeneric}. Since there it has also been found that
the Manning limit need no longer act as a boundary in the multivalent
case, see Fig.~\ref{pic:press_v123}, it would be very interesting to
perform experiments on those systems with, e.g., divalent ions.
Observe that the experimental value of $\oscoeff$ for the highly
charged dilute system 2 is already at the Manning limit.

Since the experimentally determined osmotic coefficient appears to be
smaller even than the molecular dynamics results, this indicates
effects to be relevant which go beyond the model used for
simulation. Most obvious candidates for this are the neglect of
additional chemical interactions between the ions and the
polyelectrolyte as well as solvation effects, i.e., interactions
between the ions or the polyelectrolyte with the water molecules from
the solution. It is for instance demonstrated in Ref.~\cite{blaul00a}
that the osmotic coefficient also depends on whether on uses chlorine
or iodine counterions. While one could certainly account for the
different radii of these ions when computing the distance of closest
approach entering the PB equation, the implications of the different
hydration energies is much less obvious to incorporate and in
principle requires very expensive all-atom simulations. See also the
discussion in Ref. \cite{deserno01a}, which includes the effects of
small excess salt to the osmotic coefficient.

Finally it should be noted that the two different methods of analyzing
the molecular dynamics data lead to very similar results. While this
fact is not particularly useful in a simulation, its consequences for
the analysis of experimental data seem promising. In
Ref.~\cite{guilleaume00a} poly(p-phenylene) solutions are investigated
by means of small angle X-ray scattering, which due to the rod-like
geometry turns out to be sensitive to the radial ion distribution
function. Therefore, the measured structure factors can be related to
PB distribution functions with effective values of $\xi$, $\gamma$
etc.  Fitting the scattering intensity and thereby determining those
values permits in principle a measurement of the osmotic coefficient
along the lines already described above. What makes this approach so
attractive is that it could ideally complement osmometry. If the
density becomes too large, the latter suffers from severe problems
with counter diffusion, but it is exactly this high density which
meets the requirements of good contrast necessary for scattering.

A similar fitting procedure has been applied to the measured data from
small angle X-ray scattering experiments on the systems 3 and 4; see
also Ref.~\cite{guilleaume00a}. From a fit to the structure factor the
radius $r_0^\star$ of the rod has been determined.  The integration
constant $\gamma^\ast$ and the Manning parameter $R_\rM^\ast$ the
follow from the PB theory.  The obtained values are listed in
Table~\ref{tab:ppp_fit}.  The resulting Manning radius is surprisingly
close to the one determined by simulation, particularly for the system
with $\xi=3.32$. The osmotic coefficients constructed from the PB
formula are less accurate, 10-20\% too large.


\section{The importance of correlations}\label{sec:meascorrel}

We have seen that the nonlinear PB equation suffers from systematic deviations
in strongly coupled or dense systems. It underestimates the extent of
counterion condensation and at the same time overestimates the osmotic
coefficient. The common reason for both problems is the neglect of
correlations: More elaborate theories
\cite{gonzalestovar85a,das95a,barbosa00a} which try to incorporate
correlations reproduce the trends of the simulations, because they
favor an increase in density close to the macroion surface, thereby
leading to a stronger condensation and a concomitant drop in the
boundary density and thus pressure.

On the one hand there is general agreement about the presence of
correlations being liable for the failure of PB theory. On the other
hand, this insight does not shed any light onto the question, which
kind of correlations are important. In the following we investigate a
simple pair-correlation function based on a one-rod property.

If the surface charge density of the rod is high, a fairly large
number of counterions will stay within a condensed layer of small
radial extent.  Addition of salt will further increase the ionic
density in this layer. It is clear that beyond a certain point (high
$\lambda$, high $\ell_\rB$ or enough salt) the ions will no longer
distribute independently of each other but get locally
correlated. This effect will now be measured for a DNA-like system,
because for those systems correlation effects are assumed to play a
prominent role. The rod radius was chosen to be 7.86 \AA ($1.85
\sigma$) and the ion diameter is correspondingly 4.25 \AA ($1
\sigma$).  In addition 0.5 mol/l of a 2:2 salt has been added in
excess to the divalent counterions. More information about theses
systems can be found in Table~\ref{tab:DNA_param}.

The first issue is to define the innermost layer. For this, a distance
from the rod is chosen which contains many counterions but virtually
no co-ions. A distance of roughly $11.5\,\text{\AA}$ from the rod axis
turned out to be suitable. This is about a third ion diameter farther
out than the distance of closest approach. To avoid difficulties with
remaining co-ions, only the counterions within this distance are taken
into account in all what follows.  In a second step the coordinates of
those ions are radially projected onto the surface of the cylinder of
closest approach, and this surface is then rolled out to a flat plane,
see Fig.~\ref{pic:gr2d} for an illustration of this procedure.
Finally, the two-dimensional pair correlation function $g(r)$ of these
projected points is computed.

One might object that the unrolled flat plane leads one to believe
that the pair correlation function does only depend on distance, while
in reality such a rotational symmetry cannot be expected, due to the
rod curvature. However, an investigation of $g(\VECr)$ revealed that
the pair correlation function is in fact rotationally symmetric up to
essentially a distance which corresponds to half the circumference of
the cylinder of closest approach, which is roughly
$30\,\text{\AA}$. Larger distances can of course only be realized
along the rod instead of around it. A possible reason for this
surprising symmetry is that at short distances the curvature is not
yet perceptible while at large distances the particles are already
uncorrelated.

Figure~\ref{pic:DNA_2dgr} shows the measured $g(r)$ for four
differently charged systems with Manning parameter 10.5, 8.4, 6.3, and
4.2. The last value corresponds to DNA in aqueous solution.  The most
important thing to observe is that for the strongly charged systems
$g(r)$ shows definite signs of correlations. Apart from the trivial
correlation hole at small $r$ there is a distance $r_{\max}$ at which
ions are more likely to be found than under the assumption of
independent distribution.  Notice that the maximum in $g(r)$ is not
located at $r \approx \sigma$ but much farther out. Hence, its
existence is not merely an artifact of close packing of repulsive
Lennard-Jones spheres. However, if the condensed ions are assumed to
form a triangular lattice on the surface in order to maximize their
mutual repulsion, the resulting distance $d_\triangle$ is 25 - 35 \%
larger than $r_{\max}$. Together with the only weakly pronounced
oscillations in $g(r)$ this proves the correlation induced
interactions to be rather short-ranged, yet less local than a pure
hard core. In any case, the range is several times larger than the
average salt Debye length $\ell_\rD\approx0.5\,\sigma$.

Such two-dimensional pair correlation functions for mobile ions
adsorbed on charged planes have recently been investigated
theoretically in Ref.~\cite{rouzina96a}. The mean separation
$d_\varsigma=(v e_0/\varsigma)^{1/2}$ of $v$-valent ions on the
completely neutralized plane of surface charge density $\varsigma$ is
identified as the important scaling length. The main predictions for
the strongly coupled regime $\Gamma_2 = \ell_\rB v^2/d_\varsigma > 1$
are: {\em (i)}, $g(r)$ should have a single first peak at a distance
about the size $d_\varsigma$ of the correlation hole.  {\em (ii)}, the
breadth of this peak should decrease with $\Gamma_2$ while its maximum
should increase. This trend is indeed seen to be true for the
functions in Fig.~\ref{pic:DNA_2dgr}, only $d_\varsigma$ is roughly
10--20\% larger than the actual peak position. One might want to argue
that the presence of much salt entails a further increase of the layer
density on top of the usual Gouy-Chapman prediction, but the results
obtained in Ref.~\cite{rouzina96a} are claimed to be independent of
the bulk ionic strength, if the latter is smaller than the layer
density. This is here the case, even though not always by an order of
magnitude. An alternative explanation may be based on the presence of
co-ions: Although there reside very few of them within the innermost
condensed layer, there will be an appreciable amount beyond and
possibly very close to $r_\rl$. Those ions can act as ``bridges'',
they attract counterions and thereby reduce the average closest
distance between them.

Although $r_{\max}<d_\triangle$, this does not exclude a local hexatic
ordering of the counterions. But since $g(\VECr)$ is on average
rotationally symmetric, establishing signs for this requires the
investigation of a suitable 3-point correlation function. The trick is
to break the rotational symmetry, which suggests the following
procedure: $g(r)$ is proportional to the probability of finding a
particle at a distance $r$, given that there is also a particle at the
origin. Now define $g_{\!\!\!\rightarrow}(\VECr)$ to be proportional
to the probability of finding a particle at position $\VECr$, given
that there is also a particle at the origin {\em and\/} given that the
particle which is closest to the origin is situated to the right
side. This definition is further restricted to be sensitive only for
next nearest neighbors and not arbitrary other particles. Thence, it
answers the question: ``If the nearest neighbor is on the right side,
where is the next nearest neighbor?''  A scatter plot of this
observable is shown in the left part of Fig.~\ref{pic:DNA_2dg3r}. For
this, the most highly charged DNA-sized system with Manning parameter
$\xi=10.5$ has been used. Clearly visible are the two correlation
holes around the origin and the nearest neighbor as well as a tendency
of the next nearest neighbor to be on the side opposing the nearest
neighbor. All those effects are trivially explained. However, on top
of that no preferential ion accumulations in the ``hexatic
directions'' indicated as dotted lines is perceptible. The right part
of Fig.~\ref{pic:DNA_2dg3r} shows the probability distribution of
finding the next nearest neighbor at an angle $\phi$ with respect to
the line joining nearest neighbor and origin. The correlation hole
generated by the nearest neighbor is visible, but no increase in
$p(\phi)$ at $60^\circ$, $120^\circ$ or $180^\circ$. Rather, the
measured data are compatible with a fairly structureless distribution,
as indicated by the solid line.

Since the most strongly charged system does not show hexatic order, it will
certainly be absent in the other ones. However, the coupling constant
$\Gamma_2$ introduced above has a value only about $3$ in the system with
Manning parameter $\xi=10.5$. For larger values of $\Gamma_2$ one would not
only expect much stronger correlations and hexatic ordering but even
crystallization of the adsorbed counterions into a two-dimensional Wigner
crystal \cite{rouzina96a,shklovskii99a}.  More extensive investigations to
clarify the role of correlations are presently under way \cite{deserno01d}
%
\section{Summary}
Theoretical and numerical studies of stiff linear polyelectrolytes
within the framework of a cell model have been presented.  We outlined
methods to quantify condensation for general measurements of ion
distribution functions via an inflection point criterion and by using
a generalized PB fit function.  We compared our simulational results
to predictions from mean-field PB theory, and found excellent
agreement for weakly charged systems.  The enhanced counterion
condensation caused by ionic correlations and its dependence on
parameters such as density, Bjerrum length, valence and ionic strength
was measured, and effects which qualitatively go beyond mean-field
theory, e.g., charge reversal and attractive interactions between
like-charged macroions have been found.  For systems with a high salt
content, integral equation theories have proved to give a good
qualitative and quantitative description of such systems
\cite{gonzalestovar85a,deserno01b}

For the osmotic pressure measurements we displayed for a large density
range that the measured pressure stays close to the Manning limiting
law prediction due to a fortunate error cancellation of neglected
ionic correlations and density effects which point in different
directions. We showed for the specific example of a
ppp-polyelectrolyte what differences to a PB description can be
expected for both, the ionic distribution and the osmotic pressure. We
finally analyzed some correlations for a DNA-like system, and showed
that they show signs of a strongly correlated liquid, as has been
advocated recently \cite{rouzina96a,shklovskii99a}.

To assess these theoretical ideas, computer simulations will certainly
be indispensable, since they are presently the only practicable way of
obtaining sufficiently detailed information on ionic distributions and
correlation functions. The numerical investigations presented in this
work and the correlation analysis of the obtained data are a further
step in this direction, however a more detailed analysis is presently
under way \cite{deserno01d}.
%
\vspace*{-0.2cm}
\section{Acknowledegments}
We wish to thank M. Ballauff, M. Barbosa J. Blaul, B.  Guilleaume,
F. Jimenez-Angeles, K. Kremer, M. Lozada-Cassou and S. May for various
contributions to this work. In addition we acknowledge a large
computer time grant hkf06 from NIC J\"ulich and financial support by
the German Science foundation.


\begin{appendix}


\begin{center}
\LARGE \bf Appendix:  
\end{center}

\section{Defining and computing the pressure}\label{ssec:defpress}

The pressure for the primitive cell model is a nontrivial
variable to compute for two reasons: First, the long-range electrostatics has
to be properly taken into account, and secondly, the system is inherently
anisotropic. Hence, the relevant observable is the stress tensor.

For isotropic systems, the thermodynamic definition of the pressure as
the derivative of the free energy with respect to volume leads to the
general equation
\begin{equation}
  p\,V \; = \; N k_\rB T - V\left\langle\frac{\partial U}{\partial V}\right\rangle,
  \label{eq:TD_pressure}
\end{equation}
where the angular brackets $\langle\cdots\rangle$ denote a canonical
average.  For the case of short-range interactions the contribution
from $U$ can be further simplified by using
\begin{equation}
  \frac{\partial U}{\partial V} \; = \;
  \sum_i \frac{\partial U}{\partial \VECr_i}\cdot\frac{\partial
    \VECr_i}{\partial V} \; = \;
  \sum_i (-\VECF_i)\cdot\frac{\VECr_i}{3V}.
\end{equation}
Substituting this into the pressure equation~(\ref{eq:TD_pressure})
and using $\VECF_{ij}=-\VECF_{ji}$ gives
\begin{equation}
  p\,V \; = \;  N k_\rB T
  -\frac{1}{3}\,\sum_{i<j} \left\langle\VECr_{ij}\cdot\VECF_{ij}\right\rangle
  \label{eq:p_eq1}
\end{equation}
The electrostatic contribution to the pressure can be derived
\cite{hummer98a} by realizing that the electrostatic energy $U^\rC$ is
a homogeneous function of volume with degree $-1/3$, and hence by
Euler's theorem $\partial U^\rC/\partial V = -U^\rC/3V$.  Adding this
to Eqn.~(\ref{eq:p_eq1}) finally gives the desired pressure equation
\begin{equation}
  p\,V \; = \;  N k_\rB T
  -\frac{1}{3}\,\sum_{i<j} \left\langle\VECr_{ij}\cdot\VECF_{ij}^{\,\rsr}\right\rangle
  + \frac{1}{3}\,\left\langle U^\rC \right\rangle,
  \label{eq:p_eq2}
\end{equation}
where $\VECF_{ij}^{\,\rsr}$ are the short-range pair forces.  Note
that for a repulsive hard core the virial contribution is positive,
while the electrostatic contribution is one third of the energy
density, which for neutral systems is usually negative.

For anisotropic systems the stress tensor is the relevant observable
to compute. Whereas the ideal gas contribution to the pressure still
remains isotropic, the scalar product in the virial is replaced by the
tensor product ``$\otimes$''.  For the case of electrostatic
interactions the derivation is more complicated and will not be
presented here. The reader is referred to Ref.~\cite{essmann95a}.  The
result is that within the framework of Ewald techniques the
electrostatic contribution to the stress tensor, $\sfp^\rC$, can be
decomposed additively into a real space contribution $\sfp^{(r)}$ and
a Fourier space contribution $\sfp^{(k)}$. They are given by
\begin{eqnarray}
  \sfp^{(r)} V & = & \frac{1}{2} \sum_{i,j} q_i\,q_j\!\! \sum_{\VECm\in\ZZ^{3}}^{\prime}
  \bigg[\frac{2\alpha}{\sqrt{\pi}}\,\re^{-\alpha^{2}\VECr_{ij\VECm}^{2}}
  + \, \frac{\erfc(\alpha|\VECr_{ij\VECm}|)}{|\VECr_{ij\VECm}|}
  \bigg] \frac{\VECr_{ij\VECm}\otimes\VECr_{ij\VECm}}{|\VECr_{ij\VECm}|^{2}}\qquad
  {} \label{eq:pCr} \\
  \sfp^{(k)} V & = & \frac{1}{2V}\sum_{\VECk\ne\VECzero}
  \frac{4\pi}{k^{2}}\,\re^{-k^{2}/4\alpha^{2}}
  \big|\tilde{\rho}(\VECk)\big|^2 \bigg[\II -
  2\,\Big(1+\frac{k^{2}}{4\alpha^{2}}\Big)\frac{\VECk\otimes\VECk}{k^2}\bigg].
  \label{eq:pCk}
\end{eqnarray}
Here, $\VECr_{ij\VECm}=\VECr_i-\VECr_j+\VECm L_\rb$, and the canonical
average has not been denoted explicitly. For details of the notation
and the Ewald techniques in general, the reader is referred to the
literature \cite{deserno98a,deserno98b}. Note that the connection to
the isotropic case requires the trace of the pressure tensor to be
equal to the mean electrostatic energy density, i.e., $\rTr\,\sfp^\rC
\; = \; U^\rC/V$.

After computing the electrostatic stress tensor, it can be rotated such that
the rod points along the $z$-axis. The $xx$- and $yy$-components of the new
tensor then give the pressure perpendicular to the rod, while the
$zz$-component is the contribution parallel to the rod. 

\end{appendix}




\clearpage

\centerline{\Large \bf List of Tables:}
\begin{table}[h]
  \begin{center}
    \small
    \begin{tabular}{cd{2}cd{2}d{2}d{3}d{2}}
      \toprule
      system & 
      \multicolumn{1}{c}{$c$ [g/l]} & 
      \multicolumn{1}{c}{$T$ [$^\circ$C]} & 
      \multicolumn{1}{c}{$\xi$} &
      \multicolumn{1}{c}{$\sigma$ [\AA]} & 
      \multicolumn{1}{c}{$d_\rca$ [\AA]} & 
      \multicolumn{1}{c}{$R$ [\AA]} \\
      \midrule
      1 &  1.5  & 40 & 3.4  & 7.9 & 4.4 & 239.3 \\
      2 &  1.5  & 40 & 6.8  & 8.1 & 4.4 & 292.0 \\
      3 & 19.95 & 25 & 3.32 & 7.9 & 4.4 & 65.6  \\
      4 & 17.99 & 25 & 6.64 & 8.1 & 4.4 & 84.3  \\
      \bottomrule
    \end{tabular}
  \end{center}
  \caption{Mapping for four poly(p-phenylene) systems. Tabulated are
    polyelectrolyte concentration $c$, temperature $T$, Manning parameter $\xi$,
    ion diameter $\sigma$, distance of closest approach $d_\rca$ between ions
    and the rod and cell radius $R$ corresponding to the given
    concentration. The experiments have been performed under salt-free aqueous
    conditions with monovalent counterions \cite{guilleaume00a}.
}\label{tab:ppp}
\end{table}
%
\begin{table}
  \begin{center}
    \small
    \begin{tabular}{cd{4}d{4}d{7}d{3}d{6}}
      \toprule
      system & 
      \multicolumn{1}{c}{$\oscoeff_{\,\rM}$} &
      \multicolumn{1}{c}{$\oscoeff_{\,\PB}$} & 
      \multicolumn{1}{c}{$\oscoeff_{\,\rMD,1}$} &
      \multicolumn{1}{c}{$\oscoeff_{\,\rMD,2}$} &
      \multicolumn{1}{c}{$\oscoeff_{\,\exp}$} \\
      \midrule
      1 & 0.1471 & 0.2128 & 0.2005(57) & 0.201 & 0.185(15) \\
      2 & 0.0735 & 0.1073 & 0.1003(60) & 0.102 & 0.073(15) \\
      3 & 0.1506 & 0.2848 & 0.2424(57) & 0.260 & \multicolumn{1}{c}{---} \\
      4 & 0.0753 & 0.1428 & 0.1215(56) & 0.129 & \multicolumn{1}{c}{---} \\
      \bottomrule
    \end{tabular}
  \end{center}
  \vspace*{-0.5cm}
  \caption{Osmotic coefficient of the four poly(p-phenylene) systems from
    Table~\ref{tab:ppp}. Shown are the Manning limiting law (M), the
    PB prediction, the
    simulation result using the stress tensor (MD,1), the simulation result
    using a PB fit as described in Sec.~\ref{ssec:fitPB} (MD,2)
    and results from measurements on this system using osmometry \cite{blaul00a}}
\label{tab:ppp_oscoeff} 
\end{table}\clearpage
%
\begin{table}
  \begin{center}
    \small
    \begin{tabular}{cd{2}d{2}d{2}d{3}d{3}d{3}}
      \toprule
      system & 
      \multicolumn{1}{c}{$\xi$} &
      \multicolumn{1}{c}{$R$ [\AA]} & 
      \multicolumn{1}{c}{$r_0^\star$ [\AA]} &
      \multicolumn{1}{c}{$R_\rM^\ast$ [\AA]} &
      \multicolumn{1}{c}{$\gamma^\ast$} &
      \multicolumn{1}{c}{$\oscoeff^\ast$} \\
      \midrule
      3 & 3.32 & 65.6 & 8.2 & 28.19 & 0.957 & 0.289 \\
      4 & 6.64 & 84.3 & 9.9 & 39.14 & 1.014 & 0.153 \\
      \bottomrule
    \end{tabular}
  \end{center}
  \vspace*{-0.5cm}
  \caption{Fit evaluation of small angle X-ray scattering experiments on the
    systems 3 and 4 \cite{guilleaume00a}. $\xi$ is the Manning
    parameter, $R$ the cell radius implied by the polyelectrolyte
    concentration, and $r_0^\star$ the rod radius, which had been the only
    fitting parameter. From this, the integration constant $\gamma^\ast$ and
    the Manning radius $R_\rM^\ast$ is determined from the solution of the
    PB equation.  $\oscoeff^\ast=(1+{\gamma^\ast}^2)/2\xi$ is
    the attempt to predict the osmotic coefficient of the solution with the
    help of the PB formula.}\label{tab:ppp_fit} 
\end{table}

\begin{table}
  \begin{center}
    \small
    \begin{tabular}{lccc}
      \toprule
      parameter & symbol & value & value in LJ units \\
      \midrule
      ion diameter & $\sigma$ & 4.25 \AA & $\sigma$ \\
      ion valence & $v$ & 2 & 2 \\
      rod radius & $r_0$ & 7.86 \AA & $1.85\;\sigma$ \\
      line charge density (DNA) & $\lambda$ & $e_0\big/1.7\,\text{\AA}$ & $2.5\;e_0/\sigma$ \\
      Bjerrum length (water) & $\ell_\rB$ & 7.14 \AA & $1.68\;\sigma$ \\
      Manning parameter & $\xi$ & 4.2 & 4.2 \\
      box size & $L_\rb$ & 122.4 \AA & $28.8\,\sigma$ \\
      corresponding cell radius & $R$ & 69.1 \AA & $16.2\;\sigma$ \\
      temperature & $T$ & 298 K & $\epsilon/k_\rB$ \\
      \bottomrule
    \end{tabular}
  \end{center}
  \caption{Simulation parameters for a DNA-like system.}\label{tab:DNA_param}
\end{table}
\clearpage \newpage

\centerline{\Large \bf List of Figures}

\begin{figure}
\begin{center} 
\setlength{\unitlength}{0.1bp}
\begin{picture}(3960,1620)(0,0)
\special{psfile=pictures/mono_dens_scan llx=0 lly=0 urx=792 ury=378 rwi=7920}
\put(3342,544){\makebox(0,0)[l]{$\ell_\rB/\sigma = 3$}}
\put(3139,50){\makebox(0,0){$r/r_0$}}
\put(3872,200){\makebox(0,0){100}}
\put(3644,200){\makebox(0,0){50}}
\put(3342,200){\makebox(0,0){20}}
\put(3113,200){\makebox(0,0){10}}
\put(2885,200){\makebox(0,0){5}}
\put(2582,200){\makebox(0,0){2}}
\put(2354,200){\makebox(0,0){1}}
\put(2269,1520){\makebox(0,0)[r]{1.0}}
\put(2269,1276){\makebox(0,0)[r]{0.8}}
\put(2269,1032){\makebox(0,0)[r]{0.6}}
\put(2269,788){\makebox(0,0)[r]{0.4}}
\put(2269,544){\makebox(0,0)[r]{0.2}}
\put(2269,300){\makebox(0,0)[r]{0.0}}
\put(1366,544){\makebox(0,0)[l]{$\ell_\rB/\sigma = 2$}}
\put(1165,50){\makebox(0,0){$r/r_0$}}
\put(50,910){%
\special{ps: gsave currentpoint currentpoint translate
270 rotate neg exch neg exch translate}%
\makebox(0,0)[b]{\shortstack{$P(r)$}}%
\special{ps: currentpoint grestore moveto}%
}
\put(1894,200){\makebox(0,0){100}}
\put(1667,200){\makebox(0,0){50}}
\put(1366,200){\makebox(0,0){20}}
\put(1139,200){\makebox(0,0){10}}
\put(912,200){\makebox(0,0){5}}
\put(612,200){\makebox(0,0){2}}
\put(385,200){\makebox(0,0){1}}
\put(300,1520){\makebox(0,0)[r]{1.0}}
\put(300,1276){\makebox(0,0)[r]{0.8}}
\put(300,1032){\makebox(0,0)[r]{0.6}}
\put(300,788){\makebox(0,0)[r]{0.4}}
\put(300,544){\makebox(0,0)[r]{0.2}}
\put(300,300){\makebox(0,0)[r]{0.0}}
\end{picture}    \end{center}
  \caption{Counterion distribution functions $P(r)$, as defined in
    Eqn.~(\ref{probdistri}), for various simulated densities. Note that an
    increasing cell radius corresponds to functions extending towards larger
    values of $r$.  The heavy dots mark the points of inflection in $P$ as a
    function of $\ln(r)$, while the crosses mark the positions at which those
    points would be located on the corresponding PB distribution functions.
    For the sake of clarity such a PB distribution is only plotted for the
    system with lowest density, i.e., $R=123.85\,\sigma$ (dotted
    line).}\label{pic:mono_dens_scan}

\end{figure}
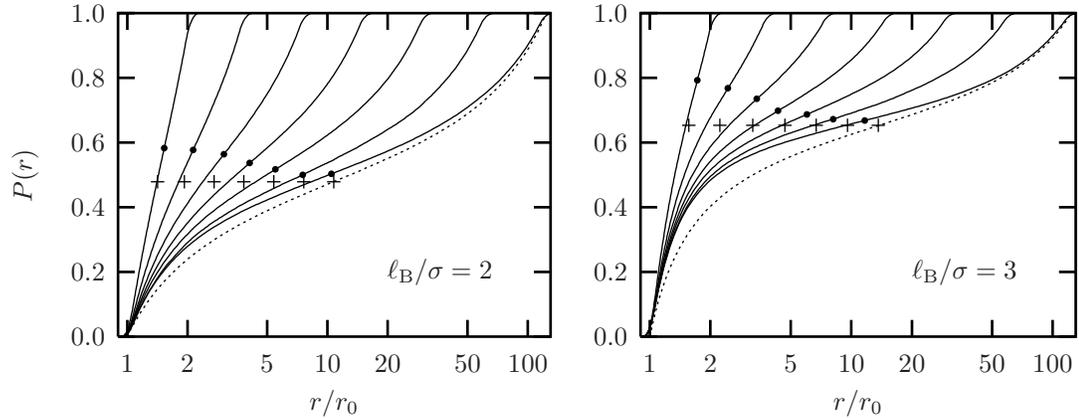

\begin{figure}
\begin{center} 
\setlength{\unitlength}{0.1bp}
\begin{picture}(3600,2160)(0,0)
\special{psfile=pictures/mono_dens_scan_Bj1 llx=0 lly=0 urx=720 ury=504 rwi=7200}
\put(1925,50){\makebox(0,0){$r/r_0$}}
\put(50,1180){%
\special{ps: gsave currentpoint currentpoint translate
270 rotate neg exch neg exch translate}%
\makebox(0,0)[b]{\shortstack{$P(r)$}}%
\special{ps: currentpoint grestore moveto}%
}
\put(3334,200){\makebox(0,0){100}}
\put(2895,200){\makebox(0,0){50}}
\put(2314,200){\makebox(0,0){20}}
\put(1875,200){\makebox(0,0){10}}
\put(1436,200){\makebox(0,0){5}}
\put(856,200){\makebox(0,0){2}}
\put(417,200){\makebox(0,0){1}}
\put(300,2060){\makebox(0,0)[r]{1.0}}
\put(300,1708){\makebox(0,0)[r]{0.8}}
\put(300,1356){\makebox(0,0)[r]{0.6}}
\put(300,1004){\makebox(0,0)[r]{0.4}}
\put(300,652){\makebox(0,0)[r]{0.2}}
\put(300,300){\makebox(0,0)[r]{0.0}}
\end{picture} \end{center}
  \caption{Counterion distribution functions $P(r)$ (solid lines) for seven
    systems with the same dimensions as the ones in
    Fig.~\ref{pic:mono_dens_scan}, 
    but with a Bjerrum length $\ell_\rB/\sigma = 1$. Since the resulting
    Manning parameter $\xi=0.959<1$, counterion condensation is not expected
    to occur. This is borne out by the observation that the functions are
    convex up already at $r=r_0$. In these weakly charged systems the
    predictions of PB theory (dotted lines) are excellent and
    can hardly be distinguished from the simulation
    results.}\label{pic:mono_dens_scan_Bj1}

\end{figure}
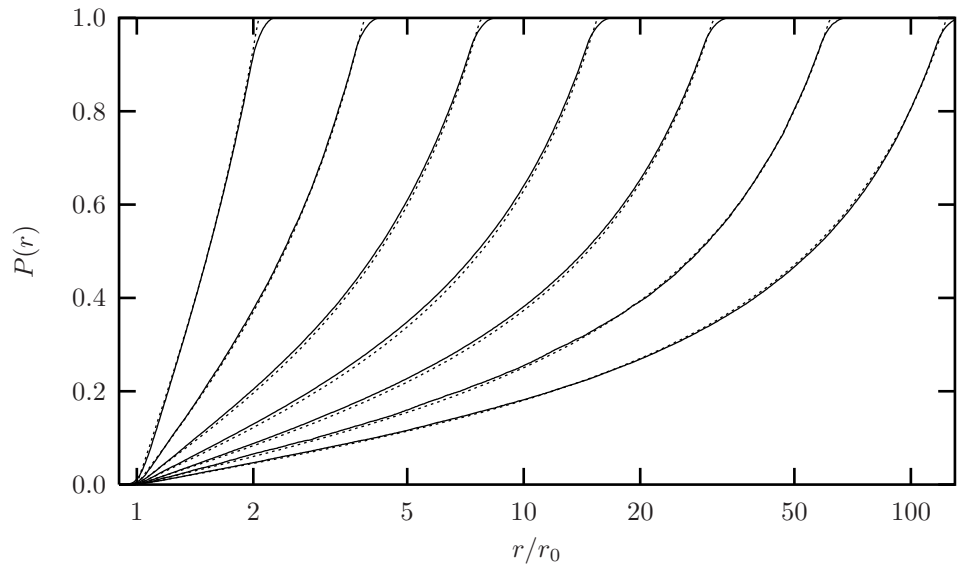

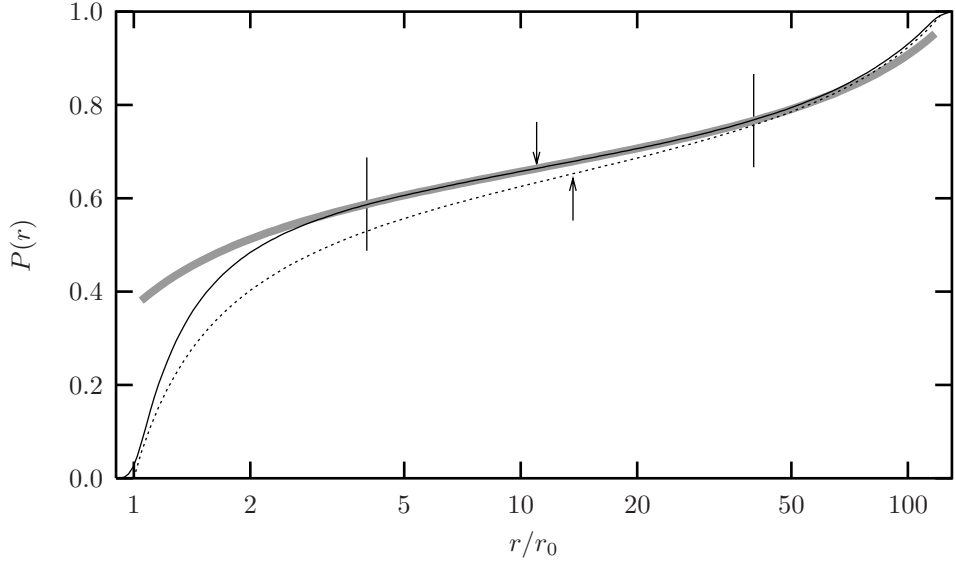
\begin{figure}
\begin{center}  
\setlength{\unitlength}{0.1bp}
\begin{picture}(3600,2160)(0,0)
\special{psfile=pictures/mono_PB_fit llx=0 lly=0 urx=720 ury=504 rwi=7200}
\put(1925,50){\makebox(0,0){$r/r_0$}}
\put(50,1180){%
\special{ps: gsave currentpoint currentpoint translate
270 rotate neg exch neg exch translate}%
\makebox(0,0)[b]{\shortstack{$P(r)$}}%
\special{ps: currentpoint grestore moveto}%
}
\put(3334,200){\makebox(0,0){100}}
\put(2895,200){\makebox(0,0){50}}
\put(2314,200){\makebox(0,0){20}}
\put(1875,200){\makebox(0,0){10}}
\put(1436,200){\makebox(0,0){5}}
\put(856,200){\makebox(0,0){2}}
\put(417,200){\makebox(0,0){1}}
\put(300,2060){\makebox(0,0)[r]{1.0}}
\put(300,1708){\makebox(0,0)[r]{0.8}}
\put(300,1356){\makebox(0,0)[r]{0.6}}
\put(300,1004){\makebox(0,0)[r]{0.4}}
\put(300,652){\makebox(0,0)[r]{0.2}}
\put(300,300){\makebox(0,0)[r]{0.0}}
\end{picture}   \end{center}
  \caption{The functional form in Eqn.~(\ref{eq:P_fit}) has been fitted to
    the measured distribution function (solid line) of the system with
    $R=123.85\,\sigma$ and $\ell_\rB=3\,\sigma$ within the range
    $[4\,r_0;40\,r_0]$ between the two vertical bars.  The $\uparrow$-arrow
    indicates the inflection point of the PB distribution
    (dotted line), while the $\downarrow$-arrow indicates the corresponding
    point $(R_\rM^\star;f_\rM^\star) = (11.0\,\sigma;0.664)$ of the fit
    (gray stripe). Note that $R_\rM^\star<R_\rM$ although $f_\rM^\star>f_\rM$.
    The result of the fit is $\xi^\star=2.97$, $\gamma^\star=0.457$ and
    $r_0^\star=0.579\,\sigma$.}\label{pic:mono_PB_fit}

\end{figure}

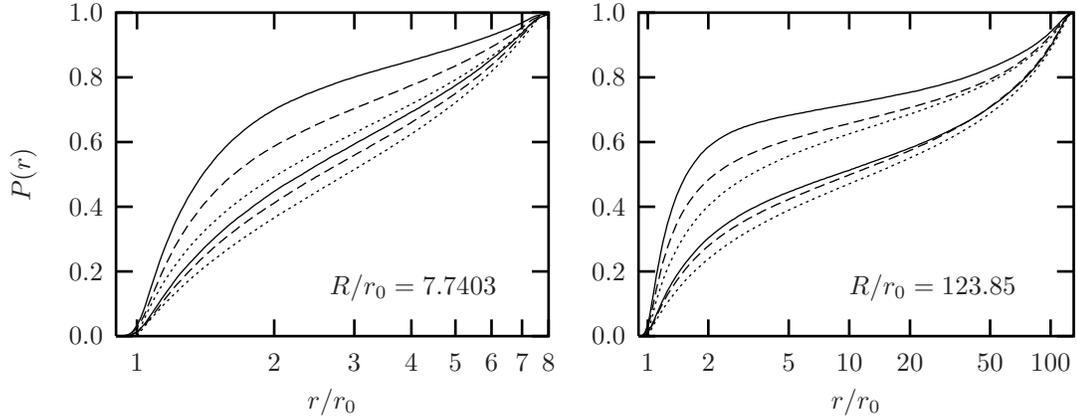
\begin{figure}
\begin{center}  
\setlength{\unitlength}{0.1bp}
\begin{picture}(3960,1620)(0,0)
\special{psfile=pictures/mono_bi_tri llx=0 lly=0 urx=792 ury=378 rwi=7920}
\put(3113,483){\makebox(0,0)[l]{$R/r_0 = 123.85$}}
\put(3139,50){\makebox(0,0){$r/r_0$}}
\put(3872,200){\makebox(0,0){100}}
\put(3644,200){\makebox(0,0){50}}
\put(3342,200){\makebox(0,0){20}}
\put(3113,200){\makebox(0,0){10}}
\put(2885,200){\makebox(0,0){5}}
\put(2582,200){\makebox(0,0){2}}
\put(2354,200){\makebox(0,0){1}}
\put(2269,1520){\makebox(0,0)[r]{1.0}}
\put(2269,1276){\makebox(0,0)[r]{0.8}}
\put(2269,1032){\makebox(0,0)[r]{0.6}}
\put(2269,788){\makebox(0,0)[r]{0.4}}
\put(2269,544){\makebox(0,0)[r]{0.2}}
\put(2269,300){\makebox(0,0)[r]{0.0}}
\put(1156,483){\makebox(0,0)[l]{$R/r_0 = 7.7403$}}
\put(1165,50){\makebox(0,0){$r/r_0$}}
\put(50,910){%
\special{ps: gsave currentpoint currentpoint translate
270 rotate neg exch neg exch translate}%
\makebox(0,0)[b]{\shortstack{$P(r)$}}%
\special{ps: currentpoint grestore moveto}%
}
\put(1980,200){\makebox(0,0){8}}
\put(1880,200){\makebox(0,0){7}}
\put(1765,200){\makebox(0,0){6}}
\put(1629,200){\makebox(0,0){5}}
\put(1463,200){\makebox(0,0){4}}
\put(1248,200){\makebox(0,0){3}}
\put(946,200){\makebox(0,0){2}}
\put(429,200){\makebox(0,0){1}}
\put(300,1520){\makebox(0,0)[r]{1.0}}
\put(300,1276){\makebox(0,0)[r]{0.8}}
\put(300,1032){\makebox(0,0)[r]{0.6}}
\put(300,788){\makebox(0,0)[r]{0.4}}
\put(300,544){\makebox(0,0)[r]{0.2}}
\put(300,300){\makebox(0,0)[r]{0.0}}
\end{picture}   \end{center}
  \caption{Counterion distribution functions $P(r)$ versus $r/r_0$. The high
    (low) density situation is shown in the left (right) frame. The lower
    three curves are for $v\,\ell_\rB/\sigma=2$, while the upper three
    correspond to $v\,\ell_\rB/\sigma=3$. The systems with multivalent
    counterions (solid lines) always show a stronger condensation than the
    complementary systems with monovalent ions (dashed lines), which
    themselves show a stronger condensation than PB theory
    (dotted lines).}\label{pic:mono_bi_tri}

\end{figure}

\begin{figure}
\begin{center} 
\setlength{\unitlength}{0.1bp}
\begin{picture}(3600,2160)(0,0)
\special{psfile=pictures/SW575859 llx=0 lly=0 urx=720 ury=504 rwi=7200}
\put(1925,50){\makebox(0,0){$r/r_0$}}
\put(50,1180){%
\special{ps: gsave currentpoint currentpoint translate
270 rotate neg exch neg exch translate}%
\makebox(0,0)[b]{\shortstack{$P(r)$}}%
\special{ps: currentpoint grestore moveto}%
}
\put(3257,200){\makebox(0,0){50}}
\put(2594,200){\makebox(0,0){20}}
\put(2092,200){\makebox(0,0){10}}
\put(1591,200){\makebox(0,0){5}}
\put(928,200){\makebox(0,0){2}}
\put(426,200){\makebox(0,0){1}}
\put(300,2060){\makebox(0,0)[r]{1.0}}
\put(300,1708){\makebox(0,0)[r]{0.8}}
\put(300,1356){\makebox(0,0)[r]{0.6}}
\put(300,1004){\makebox(0,0)[r]{0.4}}
\put(300,652){\makebox(0,0)[r]{0.2}}
\put(300,300){\makebox(0,0)[r]{0.0}}
\end{picture}   \end{center}
  \caption{Distribution functions $P(r)$ for a system 
    which has $R/\sigma=61.923$, $\ell_\rB/\sigma=2.189$,
    $\xi=2.1$ and monovalent ions. From bottom to top the number of salt
    molecules added to the simulation box of length $L_\rb=144.446\,\sigma$ is
    0, 104 and 3070, which corresponds to a salt Debye length of $\infty$,
    $22.9\,\sigma$ and $4.2\,\sigma$, respectively. The solid lines are the
    result of a simulation while the dotted lines are the predictions of
    PB theory.}\label{pic:SW575859}

\end{figure}
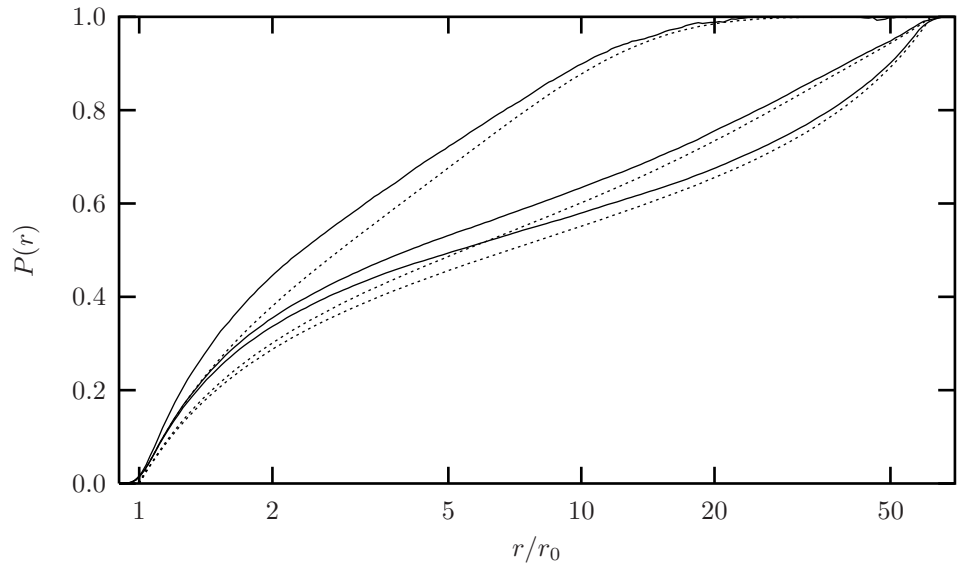

\begin{figure}
\begin{center} 
\setlength{\unitlength}{0.1bp}
\begin{picture}(3960,1620)(0,0)
\special{psfile=pictures/challenge llx=0 lly=0 urx=792 ury=378 rwi=7920}
\put(3415,717){\makebox(0,0){$r/r_0$}}
\put(2703,982){%
\special{ps: gsave currentpoint currentpoint translate
270 rotate neg exch neg exch translate}%
\makebox(0,0)[b]{\shortstack{$y(r)$}}%
\special{ps: currentpoint grestore moveto}%
}
\put(3713,467){\makebox(0,0){5}}
\put(3620,467){\makebox(0,0){4}}
\put(3501,467){\makebox(0,0){3}}
\put(3333,467){\makebox(0,0){2}}
\put(3045,467){\makebox(0,0){1}}
\put(2928,1347){\makebox(0,0)[r]{0.5}}
\put(2928,1124){\makebox(0,0)[r]{0.0}}
\put(2928,901){\makebox(0,0)[r]{-0.5}}
\put(2928,678){\makebox(0,0)[r]{-1.0}}
\put(3163,50){\makebox(0,0){$r/r_0$}}
\put(2091,910){%
\special{ps: gsave currentpoint currentpoint translate
270 rotate neg exch neg exch translate}%
\makebox(0,0)[b]{\shortstack{$n(r)\,\sigma^3$}}%
\special{ps: currentpoint grestore moveto}%
}
\put(3960,200){\makebox(0,0){17}}
\put(3672,200){\makebox(0,0){10}}
\put(3296,200){\makebox(0,0){5}}
\put(3175,200){\makebox(0,0){4}}
\put(3019,200){\makebox(0,0){3}}
\put(2799,200){\makebox(0,0){2}}
\put(2423,200){\makebox(0,0){1}}
\put(2316,1520){\makebox(0,0)[r]{0.03}}
\put(2316,1113){\makebox(0,0)[r]{0.02}}
\put(2316,707){\makebox(0,0)[r]{0.01}}
\put(2316,300){\makebox(0,0)[r]{0.00}}
\put(1103,50){\makebox(0,0){$r/r_0$}}
\put(75,910){%
\special{ps: gsave currentpoint currentpoint translate
270 rotate neg exch neg exch translate}%
\makebox(0,0)[b]{\shortstack{$P(r)$}}%
\special{ps: currentpoint grestore moveto}%
}
\put(1906,200){\makebox(0,0){17}}
\put(1616,200){\makebox(0,0){10}}
\put(1237,200){\makebox(0,0){5}}
\put(1115,200){\makebox(0,0){4}}
\put(958,200){\makebox(0,0){3}}
\put(736,200){\makebox(0,0){2}}
\put(358,200){\makebox(0,0){1}}
\put(250,1520){\makebox(0,0)[r]{1.5}}
\put(250,1113){\makebox(0,0)[r]{1.0}}
\put(250,707){\makebox(0,0)[r]{0.5}}
\put(250,300){\makebox(0,0)[r]{0.0}}
\end{picture}   \end{center}
  \caption{Charge distribution for a system characterized by
    $L_\rb/\sigma=36.112$, i.e., 60 monovalent counterions and a cell size of
    $R/\sigma =15.481$, $\ell_\rB/\sigma=4.1698$, i.e., $\xi=4$, and 1000
    molecules of a 2:2 salt within the box volume. The simulation shows a
    pronounced overcharging-effect in $P(r)$ (solid curve, left frame), in
    contrast to PB-theory (dotted curve). The charge
    oscillations can be described quite accurately by an exponentially damped
    sine function with period $1.89\,\sigma$ and decay length $0.85\,\sigma$
    (gray stripe). The densities $n_{-2}(r)$ (solid line, right frame) and
    $n_{+2}(r)$ (dotted line) of negative and positive salt ions,
    respectively, demonstrate the effect of charge layering and local charge
    reversal, and the inlay shows the dimensionless electrostatic potential
    $y(r)=\beta e_0 \psi(r)$, which is also oscillating.}\label{pic:challenge}

\end{figure}
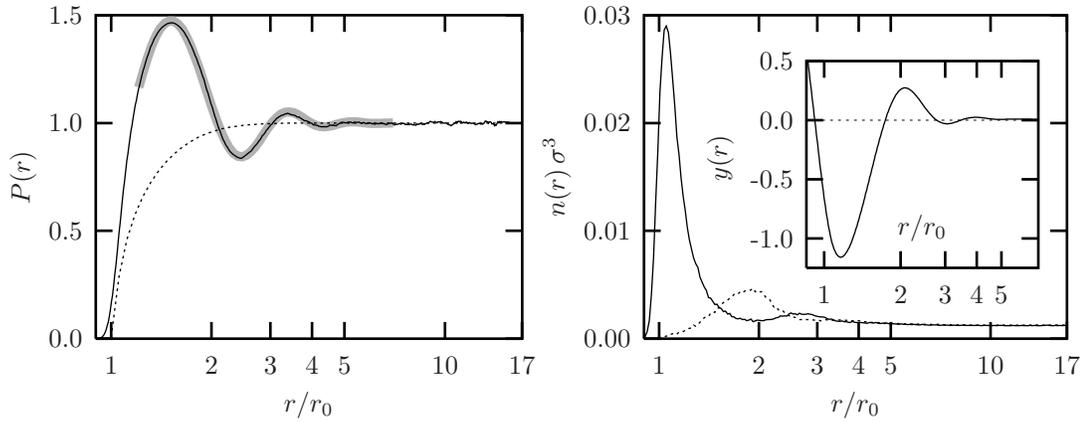

\begin{figure}
\begin{center} 
\setlength{\unitlength}{0.1bp}
\begin{picture}(3600,2160)(0,0)
\special{psfile=pictures/pressure_Bj123 llx=0 lly=0 urx=720 ury=504 rwi=7200}
\put(1925,50){\makebox(0,0){$n\,\sigma^3$}}
\put(50,1180){%
\special{ps: gsave currentpoint currentpoint translate
270 rotate neg exch neg exch translate}%
\makebox(0,0)[b]{\shortstack{osmotic coefficient $\oscoeff$}}%
\special{ps: currentpoint grestore moveto}%
}
\put(3500,200){\makebox(0,0){$10^{-1}$}}
\put(2713,200){\makebox(0,0){$10^{-2}$}}
\put(1925,200){\makebox(0,0){$10^{-3}$}}
\put(1138,200){\makebox(0,0){$10^{-4}$}}
\put(350,200){\makebox(0,0){$10^{-5}$}}
\put(300,1884){\makebox(0,0)[r]{1.0}}
\put(300,1532){\makebox(0,0)[r]{0.8}}
\put(300,1180){\makebox(0,0)[r]{0.6}}
\put(300,828){\makebox(0,0)[r]{0.4}}
\put(300,476){\makebox(0,0)[r]{0.2}}
\end{picture}   \end{center}
  \caption{Osmotic coefficient $\oscoeff$ versus $n\sigma^3$ for monovalent
    counterions. Heavy dots mark the measurements, while the solid lines are
    fits which merely serve to guide the eye. The dotted lines are the
    prediction of PB theory. From top to bottom the Bjerrum length
    $\ell_\rB/\sigma$ varies as 1,2,3. The errors in the measurement are
    roughly as big as the dot size.}\label{pic:press_Bj123}

\end{figure}
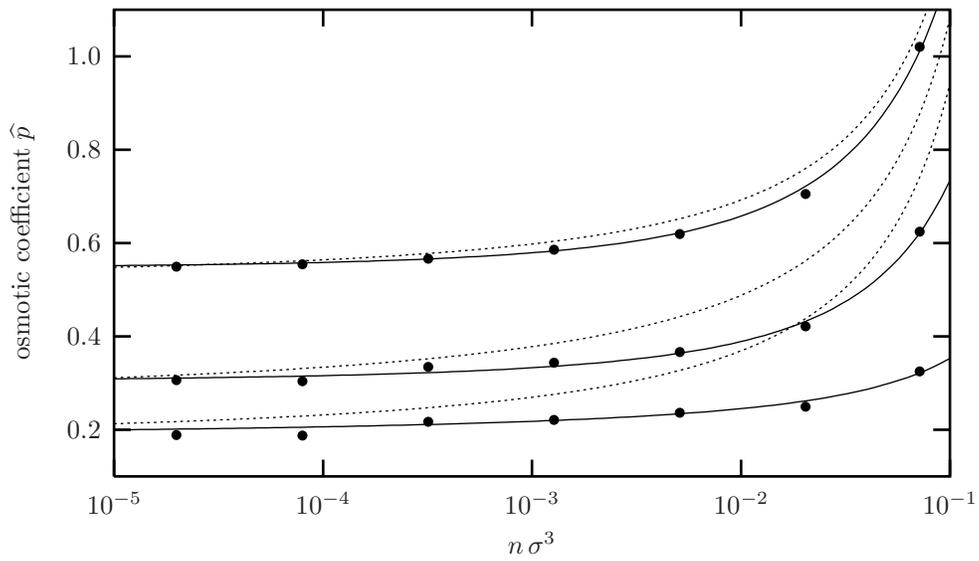

\begin{figure}
\begin{center} 
\setlength{\unitlength}{0.1bp}
\begin{picture}(3600,2160)(0,0)
\special{psfile=pictures/pressure_v123 llx=0 lly=0 urx=720 ury=504 rwi=7200}
\put(1950,50){\makebox(0,0){$n\,\sigma^3$}}
\put(50,1180){%
\special{ps: gsave currentpoint currentpoint translate
270 rotate neg exch neg exch translate}%
\makebox(0,0)[b]{\shortstack{osmotic coefficient $\oscoeff$}}%
\special{ps: currentpoint grestore moveto}%
}
\put(3324,200){\makebox(0,0){$10^{-1}$}}
\put(2739,200){\makebox(0,0){$10^{-2}$}}
\put(2154,200){\makebox(0,0){$10^{-3}$}}
\put(1570,200){\makebox(0,0){$10^{-4}$}}
\put(985,200){\makebox(0,0){$10^{-5}$}}
\put(400,200){\makebox(0,0){$10^{-6}$}}
\put(350,1995){\makebox(0,0)[r]{2.0}}
\put(350,1669){\makebox(0,0)[r]{1.5}}
\put(350,1343){\makebox(0,0)[r]{1.0}}
\put(350,1017){\makebox(0,0)[r]{0.5}}
\put(350,691){\makebox(0,0)[r]{0.0}}
\put(350,365){\makebox(0,0)[r]{-0.5}}
\end{picture}  \end{center}
  \caption{Osmotic coefficient $\oscoeff$ as a function of density $n$ for
    different valences. Heavy
    dots mark the measurements, while the solid lines are fits which merely
    serve to guide the eye. The dotted lines are the prediction of
    PB theory. From top to bottom the counter ion valence $v$
    varies like 1,2,3, which gives the same value of $\xi v$ as the curves in
    Fig.~\ref{pic:press_Bj123}.
    The errors in the measurement are roughly as big as the
    dot size.}\label{pic:press_v123}

\end{figure}
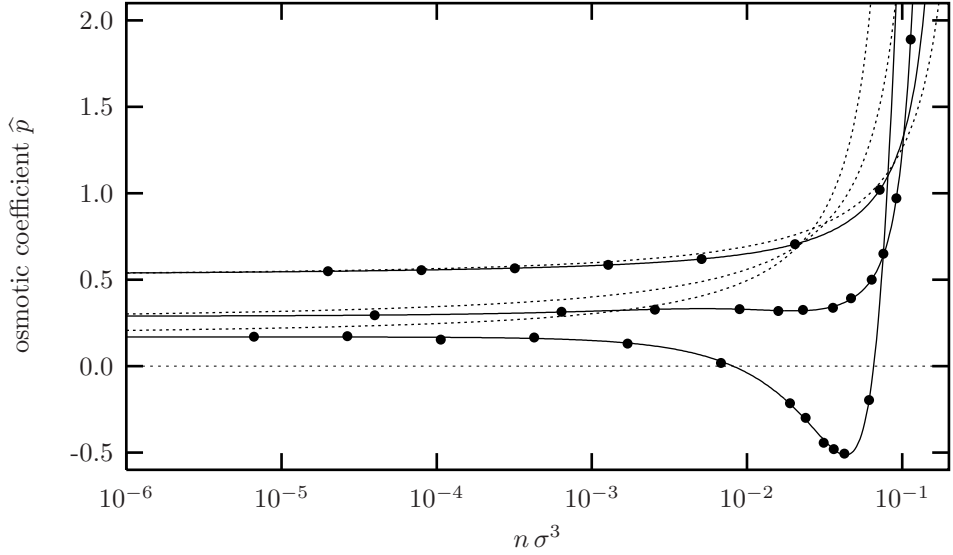

\begin{figure}
\begin{center}   
\setlength{\unitlength}{0.1bp}
\begin{picture}(3960,1620)(0,0)
\special{psfile=pictures/pressure_v23_parts llx=0 lly=0 urx=792 ury=378 rwi=7920}
\put(3139,50){\makebox(0,0){$n\,\sigma^3$}}
\put(3866,200){\makebox(0,0){$10^{-1}$}}
\put(3556,200){\makebox(0,0){$10^{-2}$}}
\put(3247,200){\makebox(0,0){$10^{-3}$}}
\put(2938,200){\makebox(0,0){$10^{-4}$}}
\put(2628,200){\makebox(0,0){$10^{-5}$}}
\put(2319,200){\makebox(0,0){$10^{-6}$}}
\put(2269,1520){\makebox(0,0)[r]{4.0}}
\put(2269,1255){\makebox(0,0)[r]{3.5}}
\put(2269,990){\makebox(0,0)[r]{3.0}}
\put(2269,724){\makebox(0,0)[r]{2.5}}
\put(2269,459){\makebox(0,0)[r]{2.0}}
\put(1165,50){\makebox(0,0){$n\,\sigma^3$}}
\put(50,910){%
\special{ps: gsave currentpoint currentpoint translate
270 rotate neg exch neg exch translate}%
\makebox(0,0)[b]{\shortstack{osmotic coefficient $\oscoeff$}}%
\special{ps: currentpoint grestore moveto}%
}
\put(1887,200){\makebox(0,0){$10^{-1}$}}
\put(1580,200){\makebox(0,0){$10^{-2}$}}
\put(1272,200){\makebox(0,0){$10^{-3}$}}
\put(965,200){\makebox(0,0){$10^{-4}$}}
\put(657,200){\makebox(0,0){$10^{-5}$}}
\put(350,200){\makebox(0,0){$10^{-6}$}}
\put(300,1520){\makebox(0,0)[r]{4.0}}
\put(300,1335){\makebox(0,0)[r]{3.5}}
\put(300,1150){\makebox(0,0)[r]{3.0}}
\put(300,965){\makebox(0,0)[r]{2.5}}
\put(300,781){\makebox(0,0)[r]{2.0}}
\put(300,596){\makebox(0,0)[r]{1.5}}
\put(300,411){\makebox(0,0)[r]{1.0}}
\end{picture}  \end{center}
  \caption{Osmotic coefficient $\oscoeff$ for the divalent (left) and
    trivalent (right) systems from Fig.~\ref{pic:press_v123}, separated into
    the non-electrostatic contribution coming from virial and ideal gas (heavy
    dots on solid lines) and negative electrostatic contribution (crosses on
    dotted lines). Again, the lines are fits which merely serve to guide the
    eye.}\label{pic:press_v23_parts}

\end{figure}
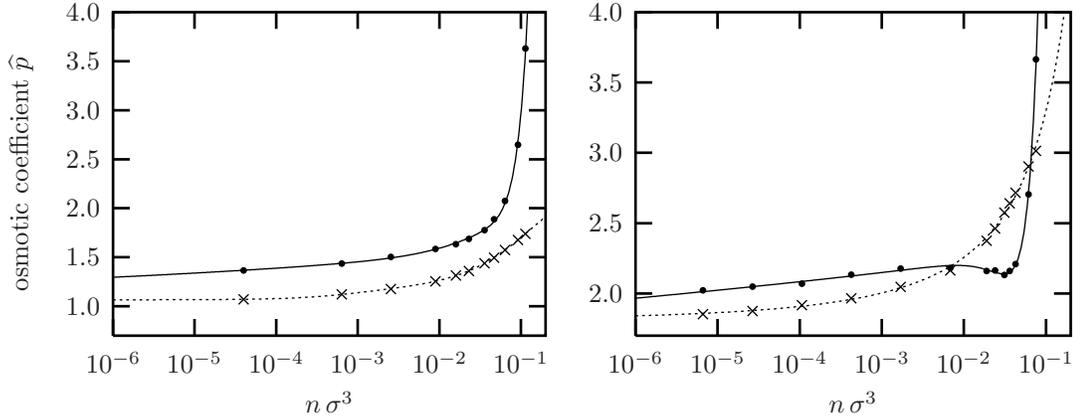

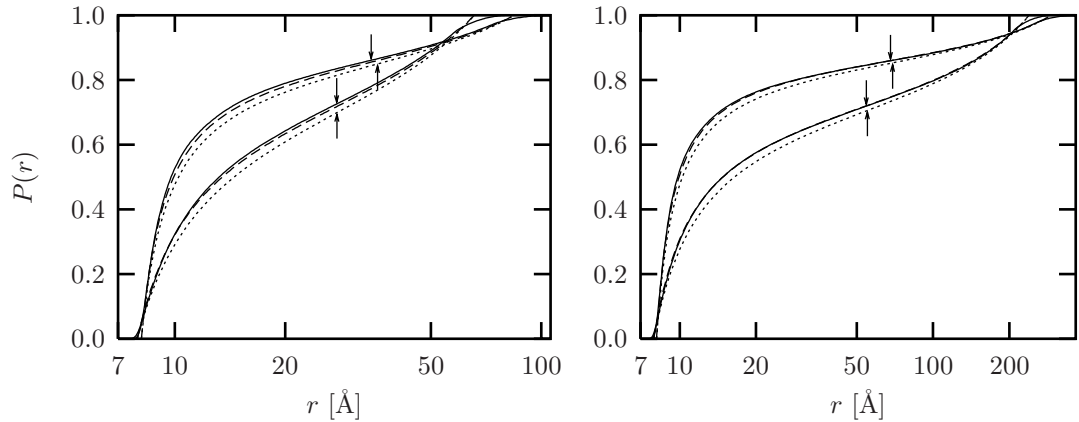
\begin{figure}
\begin{center}  
\setlength{\unitlength}{0.1bp}
\begin{picture}(3960,1620)(0,0)
\special{psfile=pictures/birgit_Pr llx=0 lly=0 urx=792 ury=378 rwi=7920}
\put(3139,50){\makebox(0,0){$r$ [\AA]}}
\put(3709,200){\makebox(0,0){200}}
\put(3421,200){\makebox(0,0){100}}
\put(3134,200){\makebox(0,0){50}}
\put(2754,200){\makebox(0,0){20}}
\put(2467,200){\makebox(0,0){10}}
\put(2319,200){\makebox(0,0){7}}
\put(2269,1520){\makebox(0,0)[r]{1.0}}
\put(2269,1276){\makebox(0,0)[r]{0.8}}
\put(2269,1032){\makebox(0,0)[r]{0.6}}
\put(2269,788){\makebox(0,0)[r]{0.4}}
\put(2269,544){\makebox(0,0)[r]{0.2}}
\put(2269,300){\makebox(0,0)[r]{0.0}}
\put(1165,50){\makebox(0,0){$r$ [\AA]}}
\put(50,910){%
\special{ps: gsave currentpoint currentpoint translate
270 rotate neg exch neg exch translate}%
\makebox(0,0)[b]{\shortstack{$P(r)$}}%
\special{ps: currentpoint grestore moveto}%
}
\put(1945,200){\makebox(0,0){100}}
\put(1529,200){\makebox(0,0){50}}
\put(980,200){\makebox(0,0){20}}
\put(564,200){\makebox(0,0){10}}
\put(350,200){\makebox(0,0){7}}
\put(300,1520){\makebox(0,0)[r]{1.0}}
\put(300,1276){\makebox(0,0)[r]{0.8}}
\put(300,1032){\makebox(0,0)[r]{0.6}}
\put(300,788){\makebox(0,0)[r]{0.4}}
\put(300,544){\makebox(0,0)[r]{0.2}}
\put(300,300){\makebox(0,0)[r]{0.0}}
\end{picture}    \end{center}
  \caption{Distribution functions for the 4 poly(p-phenylene) systems from
    Table~\ref{tab:ppp}. Left/right frame correspond to the high/low density
    systems (3,4)/(1,2), while the upper/lower set of functions correspond to
    the strongly/weakly charged systems (2,4)/(1,3) respectively. Solid lines
    are the results of simulation, dotted lines are the PB
    prediction and dashed lines are from an extended PB theory
    using the Debye-H\"uckel-hole-cavity correction \cite{barbosa00a}. The
    $\uparrow$-arrows indicate the inflection points in the PB
    distributions, while the $\downarrow$-arrows mark those points in the
    simulated distributions. The deviations of the latter from the
    PB curves at large $r$ originate from the simulation cell
    having a quadratic instead of a circular cross
    section.}\label{pic:birgit_Pr}

\end{figure}

\begin{figure}
\begin{center}  \includegraphics{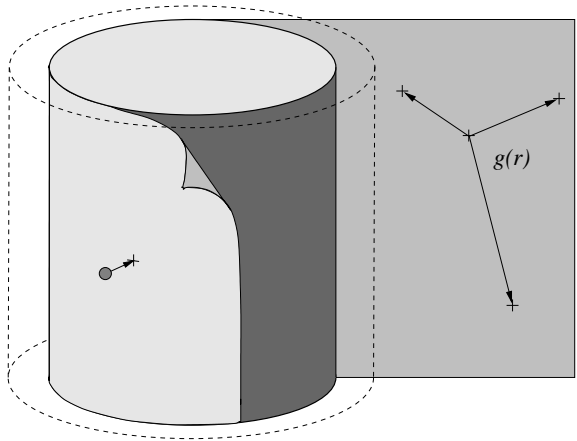}    \end{center}
    \caption{\sloppy Illustration of the computation of surface
      correlations. Counterions within a certain small distance from the rod
      constitute the innermost condensed layer. Their coordinates are radially
      projected onto the surface, which after that is unrolled to a flat
      plane. The two-dimensional pair correlation function $g(r)$ is then
      determined from the projected points.}\label{pic:gr2d}

\end{figure}

\begin{figure}
\begin{center} 
\setlength{\unitlength}{0.1bp}
\begin{picture}(3600,2160)(0,0)
\special{psfile=pictures/DNA_2dgr llx=0 lly=0 urx=720 ury=504 rwi=7200}
\put(1925,50){\makebox(0,0){$r/\sigma$}}
\put(50,1180){%
\special{ps: gsave currentpoint currentpoint translate
270 rotate neg exch neg exch translate}%
\makebox(0,0)[b]{\shortstack{two-dimensional $g(r)$}}%
\special{ps: currentpoint grestore moveto}%
}
\put(3500,200){\makebox(0,0){4}}
\put(2516,200){\makebox(0,0){3}}
\put(1531,200){\makebox(0,0){2}}
\put(547,200){\makebox(0,0){1}}
\put(300,2060){\makebox(0,0)[r]{1.2}}
\put(300,1767){\makebox(0,0)[r]{1.0}}
\put(300,1473){\makebox(0,0)[r]{0.8}}
\put(300,1180){\makebox(0,0)[r]{0.6}}
\put(300,887){\makebox(0,0)[r]{0.4}}
\put(300,593){\makebox(0,0)[r]{0.2}}
\put(300,300){\makebox(0,0)[r]{0.0}}
\end{picture}    \end{center}
  \caption{Two-dimensional pair correlation function $g(r)$ for the projection
    of counterions within a close condensed layer onto the cylinder of closest
    approach, see text and Fig.~\ref{pic:gr2d}. For the four functions the
    Manning parameter decreases as 10.5, 8.4, 6.3 and 4.2. The last value
    corresponds to DNA.}\label{pic:DNA_2dgr}

\end{figure}
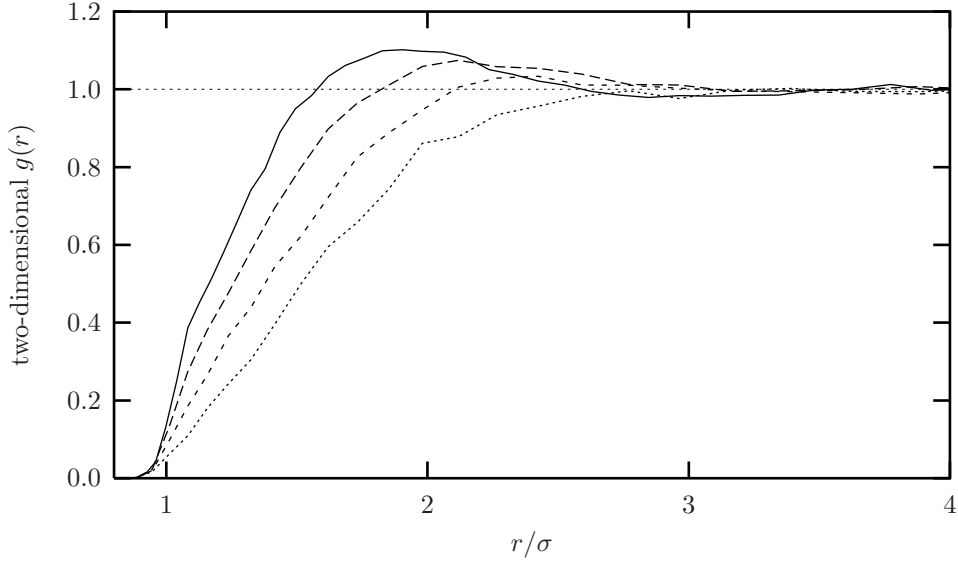

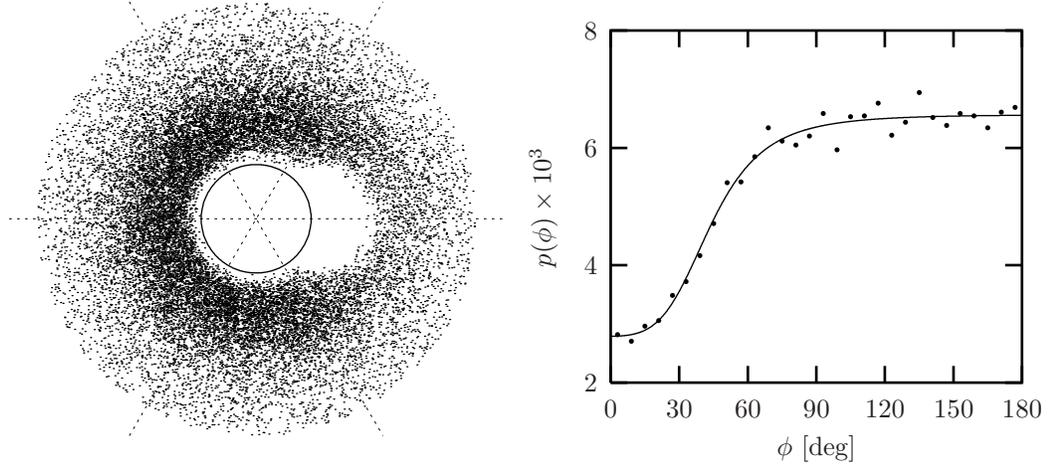
\begin{figure}
\begin{center}   
\setlength{\unitlength}{0.1bp}
\begin{picture}(3960,1620)(0,0)
\special{psfile=pictures/DNA_2dg3r llx=0 lly=0 urx=792 ury=378 rwi=7920}
\put(3185,50){\makebox(0,0){$\phi$ [deg]}}
\put(2210,964){%
\special{ps: gsave currentpoint currentpoint translate
270 rotate neg exch neg exch translate}%
\makebox(0,0)[b]{\shortstack{$p(\phi) \times 10^3$}}%
\special{ps: currentpoint grestore moveto}%
}
\put(3960,200){\makebox(0,0){180}}
\put(3702,200){\makebox(0,0){150}}
\put(3443,200){\makebox(0,0){120}}
\put(3185,200){\makebox(0,0){90}}
\put(2927,200){\makebox(0,0){60}}
\put(2668,200){\makebox(0,0){30}}
\put(2410,200){\makebox(0,0){0}}
\put(2360,1628){\makebox(0,0)[r]{8}}
\put(2360,1185){\makebox(0,0)[r]{6}}
\put(2360,743){\makebox(0,0)[r]{4}}
\put(2360,300){\makebox(0,0)[r]{2}}
\end{picture}    \end{center}
  \caption{Left part: Density plot of the 3-point correlation function
    $g_{\!\!\!\rightarrow}(\VECr)$ for the highest charged system from
    Fig.~\ref{pic:DNA_2dgr} with Manning parameter $\xi=10.5$. If an ion is
    located at the origin and its {\em nearest\/} neighbor is on the right
    side and not further away than $1.3\,\sigma$, its {\em next nearest\/}
    neighbor is shown as a dot. The circle around the origin has radius
    $\sigma$ and corresponds to the distance of closest approach imposed by
    the Lennard-Jones potential. The right frame plots the probability
    distribution of finding the next nearest neighbor at an angle $\phi$ with
    respect to a line joining nearest neighbor and origin. The solid line is a
    guide to the eye.}\label{pic:DNA_2dg3r}

\end{figure}


\end{document}